\documentclass[a4paper,fleqn,usenatbib]{mnras}

\usepackage{newtxtext,newtxmath}

\usepackage[T1]{fontenc}
\usepackage{ae,aecompl}

\usepackage{graphicx}	
\usepackage{amsmath}	
\usepackage{amssymb}	
\usepackage{xcolor}     
\usepackage{subfig}     
\usepackage{etoolbox}
\makeatletter
\patchcmd\@combinedblfloats{\box\@outputbox}{\unvbox\@outputbox}{}{%
  \errmessage{\noexpand\@combinedblfloats could not be patched}%
}%
\makeatother

\newcommand\mh{M_{\rm halo}}
\newcommand\kms{km~s$^{-1}$}
\newcommand\frise{f_{\rm rising}}
\newcommand\ffall{f_{\rm falling}}
\newcommand\gamin{\gamma_{\rm inner}}
\newcommand\gamout{\gamma_{\rm outer}}

\newcommand{\aref}[1]{\mbox{\hyperref[#1]{Appendix~\ref{#1}}}}

\title[MASSIVE VIII. - Dispersion and Environment]{The MASSIVE Survey - VIII. Stellar Velocity Dispersion Profiles and Environmental Dependence of Early-Type Galaxies} 

\author[Veale et al.]{
Melanie Veale,$^{1,2}$\thanks{E-mail: melanie.veale@berkeley.edu (MV), cpma@berkeley.edu (C-PM)}
Chung-Pei Ma,$^{1,2}$\footnotemark[1]
Jenny E. Greene,$^3$
Jens Thomas,$^4$
\newauthor
John P. Blakeslee,$^5$
Jonelle L. Walsh,$^6$
Jennifer Ito$^1$
\\
$^{1}$Department of Astronomy, University of California, Berkeley, CA 94720, USA \\
$^{2}$Department of Physics, University of California, Berkeley, CA 94720, USA \\
$^3$Department of Astrophysical Sciences, Princeton University, Princeton, NJ 08544, USA \\
$^4$Max Plank-Institute for Extraterrestrial Physics, Giessenbachstr. 1, D-85741 Garching, Germany \\
$^5$Dominion Astrophysical Observatory, NRC Herzberg Astronomy \& Astrophysics, Victoria BC V9E2E7, Canada \\
$^6$George P. and Cynthia Woods Mitchell Institute for Fundamental Physics and Astronomy, and Department of Physics and Astronomy, \\Texas A\&M University, College Station, TX 77843, USA \\
}

\date{Accepted XXX. Received YYY; in original form ZZZ}

\pubyear{2017}

\begin{document}
\label{firstpage}
\pagerange{\pageref{firstpage}--\pageref{lastpage}}
\maketitle

\begin{abstract}
  We measure the radial profiles of the stellar velocity dispersions, $\sigma(R)$, for 90 early-type galaxies (ETGs) in the MASSIVE survey, a volume-limited integral-field spectroscopic (IFS) galaxy survey targeting all northern-sky ETGs with absolute $K$-band magnitude $M_K < -25.3$ mag, or stellar mass $M_* \ga 4 \times 10^{11} M_\odot$, within 108 Mpc.
Our wide-field 107\arcsec$\times$107\arcsec\ IFS data cover radii as large as $40$~kpc, for which we quantify separately the inner ($2$~kpc) and outer ($20$~kpc) logarithmic slopes $\gamin$ and $\gamout$ of $\sigma(R)$.
While $\gamin$ is mostly negative, of the 56 galaxies with sufficient radial coverage to determine $\gamout$ we find 36\% to have rising outer dispersion profiles, 30\% to be flat within the uncertainties, and 34\% to be falling.
The fraction of galaxies with rising outer profiles increases with $M_*$ and in denser galaxy environment,
with 10 of the 11 most massive galaxies in our sample having flat or rising dispersion profiles.
The strongest environmental correlations are with local density and halo mass, but a weaker correlation with large-scale density also exists.
The average $\gamout$ is similar for brightest group galaxies, satellites, and isolated galaxies in our sample.
We find a clear positive correlation between the gradients of the outer dispersion profile and the gradients of the velocity kurtosis $h_4$. 
Altogether, our kinematic results suggest that the increasing fraction of rising dispersion profiles in the most massive ETGs are caused (at least in part) by variations in the total mass profiles rather than in the velocity anisotropy alone.
\end{abstract}




\section{Introduction}
\label{sec:introduction}

Measuring the stellar velocity dispersion profiles of elliptical galaxies is a key ingredient of estimating their dark matter content.
However, unlike spiral galaxies in which ordered rotation allows a relatively straightforward translation of rotation curves into dark matter density profiles \citep[e.g.][]{rubinetal1980}, elliptical galaxies have a complex relationship between the gravitational potential, orbital configuration of stars, and measured line-of-sight kinematics.
In addition to the measurement of dark matter, details of these relationships can also shed light on the formation and assembly history of these galaxies.

Pre-1990 long-slit measurements of the stellar velocity dispersion profiles of elliptical galaxies found most of them to decline with radius out to $\sim 10$~kpc \citep{faberetal1977,tonry1985,wilkinsonetal1986,daviesillingworth1986,franxetal1989}.
There are a few notable exceptions.
\citet{faberetal1977} found the cD galaxy in Abell~401 to have a flat dispersion profile, measuring $\sigma \sim 480$~\kms at two points (in the nucleus and at $\sim 43$~kpc).
\citet{daviesillingworth1983} found NGC~4889 and NGC~6909 to also have flat dispersion profiles out to $\sim 10$~kpc.
Rising dispersion profiles were rare.
\citet{dressler1979} found $\sigma$ for IC~1101 (BCG of Abell~2029) to rise from 375 \kms at the center to 500~\kms at 70~kpc.
\citet{carteretal1981,carteretal1985} found rising profiles for IC~2082 and the BCG of Abell~3266 (both double-nucleus galaxies) out to $\sim 30$~kpc, although
data covering smaller radial ranges
found falling or flat profiles \citep{tonry1985}.

Several long-slit studies since the 1990s have focused 
on brightest group or cluster galaxies (BGGs or BCGs), and again found mostly flat or falling dispersion profiles \citep{fisheretal1995, sembachtonry1996, carteretal1999, broughetal2007, loubseretal2008}.
More than 60 galaxies were studied in these papers, but only four galaxies were found to have rising dispersion profiles.
IC~1101 was confirmed as having a rising dispersion profile past at least 20~kpc \citep{fisheretal1995,sembachtonry1996}.
\citet{carteretal1999} found NGC~6166 to have a rising profile at $\sim 20$~kpc, although observations with a smaller field of view found flat or falling profiles \citep{tonry1985,loubseretal2008}.
The rising profile of NGC~6166 has been confirmed in \citet{kelsonetal2002} and more recently in \citet{benderetal2015}, which found it to reach the cluster dispersion of 800~\kms at a radius of 50~kpc.
This is the only study so far to confirm a smooth rise in dispersion in integrated starlight all the way to the cluster value.
The dispersion profile of NGC~3311 starts to rise around 8 kpc and appears to extrapolate smoothly to the cluster dispersion \citep{hauetal2004}, but the observations had a significant gap in radial coverage between the stellar dispersion and the cluster dispersion measured by galaxies.
\citet{loubseretal2008} and \citet{ventimigliaetal2010} confirmed the rise in NGC~3311.

However, rising dispersion profiles may be more common than this, especially at large radius.
Recently \citet{newmanetal2013} measured 7 more distant BCGs (redshifts 0.2 to 0.3) to about 30~kpc, and found all to be strikingly homogeneous in their rising dispersion profiles.
\citet{smithetal2017} also find a rising dispersion profile out to 30~kpc in the BCG of Abell 1201 at redshift 0.17.
Globular cluster (GC) measurements have identified two other galaxies with rising dispersion profiles, NGC~1407 \citep{potaetal2015b} and NGC~4486 \citep[M87;][]{cohenryzhov1997,coteetal2001,wutremaine2006}, and confirmed the rising profile in NGC~3311 \citep{richtleretal2011}.
In the case of NGC~4486, looking at starlight, \citet{sembachtonry1996} found a falling profile, but \citet{murphyetal2011,murphyetal2014} found a rise in $\sigma$ at large radius.
Other GC measurements of NGC~4486 found flat dispersion profiles to $200$~kpc \citep{straderetal2011,zhangetal2015,koetal2017}, and GC measurements of several other galaxies in a similar mass range to MASSIVE have found flat and falling dispersion profiles \citep{coteetal2003,schuberthetal2010,richtleretal2014,potaetal2013,potaetal2015}.
Inferring global population statistics from these heterogeneous results is made more difficult by the fact that different tracer populations (i.e. starlight and GCs) in general have different spatial distributions and orbital configurations, and thus different kinematics \citep[e.g.][]{potaetal2013,murphyetal2014}.

Several past and ongoing IFU surveys produce 2-dimensional maps of stellar kinematics for large samples of galaxies and provide an even more comprehensive picture of velocity dispersion profiles than long-slit observations.
The ATLAS$^{\rm 3D}$ survey \citep{emsellemetal2011} measured stellar kinematics to between $0.5 R_e$ and $1 R_e$ for a large sample of ETGs.
Data beyond $1 R_e$, however, are needed to study
the outer behaviour of the dispersion profiles.
The SLUGGS survey mapped the kinematics out to $\sim 3 R_e$ for 25 ATLAS$^{\rm 3D}$ galaxies \citep{brodieetal2014,arnoldetal2014}.  
No dispersion profiles of these galaxies have been quantified thus far, but
our visual inspections of their $\sigma(R)$ plots found only flat or falling dispersion profiles \citep{fosteretal2016}.
\citet{raskuttietal2014} measured kinematics to a few $R_e$ for 33 ETGs and found flat or declining dispersion profiles past $1 R_e$, with only the two most massive galaxies showing small increases in dispersion outside $1 R_e$.  The ongoing SAMI survey probes to between $1 R_e$ and $2 R_e$ \citep{vandesandeetal2017} and may be more likely to provide statistics of the outer dispersion profiles.

A rising dispersion profile is generally interpreted as evidence for an increasing dynamical mass-to-light ratio, but velocity anisotropy can complicate the interpretations \citep{binneymamon1982}.
Information about the full shape of the line-of-sight velocity distributions (LOSVD), in particular at least the kurtosis $h_4$, is needed to determine both the mass and the anisotropy \citep{dejonghemerritt1992, merrittsaha1993, gerhard1993}.
Gravitational lensing data often favor roughly isothermal mass profiles for ETGs \citep{treuetal2006,koopmansetal2009,augeretal2009,augeretal2010,sonnenfeldetal2013}.
There is also some evidence that mass profiles depend on galaxy mass \citep{deasonetal2012,newmanetal2013,alabietal2016} or environment \citep{newmanetal2015}, with steeper profiles at lower mass and density and shallower profiles at higher mass and density.
If mass profiles change, then dispersion profiles are also likely to change with galaxy mass or environment.
In addition, the anisotropy of a galaxy may be linked to its merger history \citep{romanowskyetal2003}, which in turn links to galaxy mass and environment, and dispersion profiles may be directly impacted by recent mergers \citep{schaueretal2014}.

In this paper we present the stellar velocity dispersion profiles, $\sigma(R)$, of 90 massive early-type galaxies in the volume-limited MASSIVE survey (\citealt{maetal2014}; Paper I).
We measure the inner and outer gradients of $\sigma(R)$ and quantity how they correlate with galaxy mass, environment, and velocity kurtosis $h_4$. 
The MASSIVE survey uses the wide-format (107\arcsec$\times$107\arcsec) Mitchell/VIRUS-P IFS to obtain 2-D kinematic maps out to radii as large as 40 kpc for the most massive local ETGs.
The survey is designed with clean sample selection criteria, targeting all ETGs with stellar mass $M_* \ga 4\times 10^{11} M_\odot$ within 108 Mpc in the northern sky, regardless of galaxy environment, size, dispersion, or other properties.
Details of our IFS data and kinematic analysis were described in \cite{vealeetal2017} (Paper V).  
A study of the relationship between MASSIVE galaxy rotation, stellar mass, and four measures of galaxy environments (group membership, halo mass $\mh$, large-scale galaxy density $\delta_g$, and local galaxy density $\nu_{10}$) was presented in \cite{vealeetal2017b} (Paper VII).  
This paper continues the stellar kinematic study of MASSIVE galaxies and focuses on the radial profiles of the stellar velocity dispersions.  

Other results from the MASSIVE survey were discussed in separate papers:
stellar population gradients (\citealt{greeneetal2015}; Paper II),
molecular gas content and kinematics (\citealt{davisetal2016}; Paper III),
X-ray halo gas properties (\citealt{gouldingetal2016}; Paper IV),
spatial distributions and kinematics of warm ionized gas (\citealt{pandyaetal2017}; Paper VI), and measurement of the black hole mass in MASSIVE galaxy NGC~1600 \citep{thomasetal2016}.

\autoref{sec:sample} of this paper summarizes our sample of galaxies and our earlier kinematic and environmental analyses.
\autoref{sec:profiles} describes how we quantify $\sigma(R)$ and presents our results for
the inner and outer power law slopes $\gamin$ and $\gamout$ of $\sigma(R)$.
\autoref{sec:kinematics} examines the relationships of the dispersion profiles
and the higher moment $h_4$, the kurtosis of the stellar velocity distribution,
and \autoref{sec:environment} examines how the dispersion profiles relate to galaxy mass and environment.
\autoref{sec:conclusions} summarizes the results and possible implications. The velocity dispersion profiles for all 90 galaxies are shown in \aref{sec:appendix-allgals}.


\section{Galaxy Sample, Properties, and Environment}
\label{sec:sample}

\begin{table*}
  \caption{Properties of MASSIVE galaxies}
  \setlength{\tabcolsep}{6pt}
\label{table:main}
\begin{tabular}{ll@{}rcccrrrrl@{\hspace{-1.3em}}rrr}
Galaxy & $M_K$ & $\log_{10} M_*$ & $\lambda_e$ & $\sigma_c$ & $\langle \sigma \rangle_e$ & $\gamma_{\rm inner}$ & $\gamma_{\rm outer}$ & $\langle h_4 \rangle$ & $h_4^\prime$ & env & $\log_{10} M_{\rm halo}$ & $1 + \delta_g$ & $\nu_{10}$ \\
 & [mag] & [$M_\odot$] &  & [km/s] & [km/s] &  &  &  &  &  & [$M_\odot$] &  & [$\overline{\nu}$] \\
(1) & (2) & (3) & (4) & (5) & (6) & (7) & (8) & (9) & (10) & (11) & (12) & (13) & (14) \\
\hline
NGC 0057 & $-25.75$ & $11.79$ & $0.022$ & $289$ & $251$ & $-0.150$ & $-0.001$ & $0.053$ & $0.017$ & I &  & $2.29$ & $4.8$ \\
NGC 0080 & $-25.66$ & $11.75$ & $0.039$ & $248$ & $222$ & $-0.073$ & $0.145$ & $0.039$ & $0.010$ & B & $14.1$ & $2.95$ & $6500$ \\
NGC 0315 & $-26.30$ & $12.03$ & $0.062$ & $348$ & $341$ & $-0.004$ & $-0.057$ & $0.052$ & $0.011$ & B & $13.5$ & $6.03$ & $270$ \\
NGC 0383 & $-25.81$ & $11.82$ & $0.252$ & $290$ & $257$ & $-0.143$ & $0.307$ & $0.012$ & $-0.019$ & S & $14.4$ & $7.24$ & $4300$ \\
NGC 0410 & $-25.90$ & $11.86$ & $0.034$ & $291$ & $247$ & $-0.121$ & $-0.276$ & $0.041$ & $-0.028$ & B & $14.4$ & $7.41$ & $3100$ \\
NGC 0499 & $-25.50$ & $11.68$ & $0.060$ & $274$ & $266$ & $-0.043$ & $-0.489$ & $0.028$ & $-0.008$ & S & $14.4$ & $7.24$ & $35000$ \\
NGC 0507 & $-25.93$ & $11.87$ & $0.049$ & $274$ & $257$ & $-0.069$ & ($0.098)$ & $0.050$ & $0.035$ & B & $14.4$ & $7.24$ & $58000$ \\
NGC 0533 & $-26.05$ & $11.92$ & $0.034$ & $280$ & $258$ & $-0.056$ & $0.068$ & $0.063$ & $0.083$ & B & $13.5$ & $4.27$ & $13$ \\
NGC 0545 & $-25.83$ & $11.83$ & $0.129$ & $249$ & $231$ & $0.011$ & ($0.266)$ & $0.074$ & $0.036$ & B(A194) & $14.5$ & $5.89$ & $13000$ \\
NGC 0547 & $-25.83$ & $11.83$ & $0.056$ & $259$ & $232$ & $-0.079$ & ($-0.006)$ & $0.035$ & $0.030$ & S(A194) & $14.5$ & $5.89$ & $14000$ \\
NGC 0665 & $-25.51$ & $11.68$ & $0.402$ & $206$ & $164$ & $-0.163$ & $-0.070$ & $-0.074$ & $-0.227$ & B & $13.7$ & $3.02$ & $56$ \\
UGC 01332 & $-25.57$ & $11.71$ & $0.037$ & $248$ & $253$ & $-0.005$ & ($0.274)$ & $0.034$ & $-0.020$ & B & $13.8$ & $3.72$ & $170$ \\
NGC 0708 & $-25.65$ & $11.75$ & $0.036$ & $206$ & $219$ & $-0.010$ & $0.273$ & $0.090$ & $0.110$ & B(A262) & $14.5$ & $5.75$ & $12000$ \\
NGC 0741 & $-26.06$ & $11.93$ & $0.040$ & $292$ & $289$ & $0.012$ & ($0.068)$ & $0.043$ & $0.069$ & B & $13.8$ & $2.88$ & $130$ \\
NGC 0777 & $-25.94$ & $11.87$ & $0.046$ & $324$ & $291$ & $-0.134$ & ($-0.090)$ & $0.051$ & $0.004$ & B & $13.5$ & $5.01$ & $76$ \\
NGC 0890 & $-25.50$ & $11.68$ & $0.101$ & $207$ & $194$ & $-0.034$ & ($-0.062)$ & $-0.006$ & $-0.002$ & I &  & $4.68$ & $1.4$ \\
NGC 0910 & $-25.33$ & $11.61$ & $0.039$ & $236$ & $219$ & $-0.178$ & $0.357$ & $0.018$ & $0.034$ & S(A347) & $14.8$ & $6.17$ & $11000$ \\
NGC 0997 & $-25.40$ & $11.64$ & $0.243$ & $267$ & $215$ & $-0.195$ & ($-0.090)$ & $0.021$ & $-0.015$ & B & $13.0$ & $2.95$ & $26$ \\
NGC 1016 & $-26.33$ & $12.05$ & $0.033$ & $286$ & $279$ & $-0.028$ & $-0.005$ & $0.027$ & $-0.001$ & B & $13.9$ & $4.79$ & $55$ \\
NGC 1060 & $-26.00$ & $11.90$ & $0.023$ & $310$ & $271$ & $-0.100$ & ($-0.068)$ & $0.055$ & $0.028$ & B & $14.0$ & $3.89$ & $2000$ \\
NGC 1132 & $-25.70$ & $11.77$ & $0.061$ & $239$ & $218$ & $-0.100$ & $0.145$ & $0.022$ & $0.015$ & B & $13.6$ & $3.39$ & $8.1$ \\
NGC 1129 & $-26.14$ & $11.96$ & $0.120$ & $241$ & $259$ & $0.036$ & $0.131$ & $0.047$ & $0.043$ & B & $14.8$ & $10.72$ & $16000$ \\
NGC 1167 & $-25.64$ & $11.74$ & $0.427$ & $188$ & $172$ & $-0.085$ & $-0.427$ & $-0.068$ & $-0.166$ & B & $13.1$ & $5.01$ & $15$ \\
NGC 1226 & $-25.51$ & $11.68$ & $0.033$ & $274$ & $229$ & $-0.146$ & $-0.173$ & $0.084$ & $0.138$ & B & $13.2$ & $3.47$ & $3.0$ \\
IC0 310 & $-25.35$ & $11.61$ & $0.085$ & $218$ & $205$ & $-0.097$ & - & $0.059$ & $-0.074$ & S(Perseus) & $14.8$ & $13.18$ & $15000$ \\
NGC 1272 & $-25.80$ & $11.81$ & $0.023$ & $285$ & $250$ & $-0.075$ & $0.118$ & $0.049$ & $0.046$ & S(Perseus) & $14.8$ & $13.49$ & $390000$ \\
UGC 02783 & $-25.44$ & $11.65$ & $0.068$ & $292$ & $266$ & $-0.163$ & ($-0.143)$ & $0.018$ & $0.015$ & B & $12.6$ & $6.31$ & $17$ \\
NGC 1453 & $-25.67$ & $11.75$ & $0.201$ & $312$ & $272$ & $-0.091$ & $0.013$ & $0.044$ & $0.005$ & B & $13.9$ & $2.29$ & $87$ \\
NGC 1497 & $-25.31$ & $11.60$ & $0.474$ & $234$ & $190$ & $-0.214$ & $0.132$ & $-0.029$ & $-0.071$ & I &  & $2.69$ & $87$ \\
NGC 1600 & $-25.99$ & $11.90$ & $0.026$ & $346$ & $293$ & $-0.082$ & $-0.047$ & $0.055$ & $0.037$ & B & $14.2$ & $6.03$ & $1200$ \\
NGC 1573 & $-25.55$ & $11.70$ & $0.040$ & $288$ & $264$ & $-0.073$ & $-0.254$ & $0.018$ & $0.015$ & B & $14.1$ & $4.07$ & $580$ \\
NGC 1684 & $-25.34$ & $11.61$ & $0.122$ & $295$ & $262$ & $-0.054$ & ($-0.054)$ & $0.018$ & $0.011$ & B & $13.7$ & $6.17$ & $1500$ \\
NGC 1700 & $-25.60$ & $11.72$ & $0.195$ & $236$ & $223$ & $-0.086$ & $-0.345$ & $-0.026$ & $-0.074$ & B & $12.7$ & $3.47$ & $23$ \\
NGC 2208 & $-25.63$ & $11.74$ & $0.062$ & $268$ & $255$ & $-0.024$ & ($0.049)$ & $-0.004$ & $-0.007$ & I &  & $2.82$ & $7.1$ \\
NGC 2256 & $-25.87$ & $11.84$ & $0.024$ & $240$ & $259$ & $0.121$ & ($-0.103)$ & $0.063$ & $0.004$ & B & $13.7$ & $2.69$ & $20$ \\
NGC 2274 & $-25.69$ & $11.76$ & $0.067$ & $288$ & $259$ & $-0.083$ & $0.133$ & $0.021$ & $-0.024$ & B & $13.3$ & $3.09$ & $110$ \\
NGC 2258 & $-25.66$ & $11.75$ & $0.036$ & $293$ & $254$ & $-0.080$ & ($0.035)$ & $0.040$ & $0.042$ & B & $12.2$ & $3.80$ & $9.5$ \\
NGC 2320 & $-25.93$ & $11.87$ & $0.235$ & $340$ & $298$ & $-0.167$ & ($0.000)$ & $0.037$ & $0.028$ & B & $14.2$ & $7.94$ & $650$ \\
UGC 03683 & $-25.52$ & $11.69$ & $0.090$ & $257$ & $257$ & $-0.051$ & ($-0.093)$ & $0.024$ & $-0.090$ & B & $13.6$ & $5.75$ & $26$ \\
NGC 2332 & $-25.39$ & $11.63$ & $0.037$ & $254$ & $224$ & $-0.121$ & $-0.123$ & $0.029$ & $-0.033$ & S & $14.2$ & $7.76$ & $1500$ \\
NGC 2340 & $-25.90$ & $11.86$ & $0.029$ & $232$ & $235$ & $-0.028$ & $0.033$ & $0.018$ & $0.004$ & S & $14.2$ & $7.76$ & $1200$ \\
UGC 03894 & $-25.58$ & $11.72$ & $0.122$ & $297$ & $255$ & $-0.142$ & $-0.107$ & $0.036$ & $0.025$ & B & $13.7$ & $1.55$ & $1.5$ \\
NGC 2418 & $-25.42$ & $11.64$ & $0.241$ & $245$ & $217$ & $-0.176$ & ($-0.236)$ & $0.041$ & $-0.029$ & I &  & $2.24$ & $1.4$ \\
NGC 2513 & $-25.52$ & $11.69$ & $0.095$ & $280$ & $253$ & $-0.071$ & $-0.127$ & $-0.004$ & $-0.025$ & B & $13.6$ & $2.34$ & $5.1$ \\
NGC 2672 & $-25.60$ & $11.72$ & $0.095$ & $273$ & $262$ & $-0.055$ & ($-0.033)$ & $0.028$ & $-0.009$ & B & $13.0$ & $1.32$ & $1.2$ \\
NGC 2693 & $-25.76$ & $11.79$ & $0.295$ & $327$ & $296$ & $-0.061$ & $0.087$ & $0.035$ & $-0.011$ & I &  & $1.70$ & $6.8$ \\
NGC 2783 & $-25.72$ & $11.78$ & $0.042$ & $252$ & $264$ & $0.034$ & $0.012$ & $0.047$ & $-0.007$ & B & $12.8$ & $3.24$ & $4.6$ \\
NGC 2832 & $-26.42$ & $12.08$ & $0.071$ & $327$ & $291$ & $-0.093$ & $0.096$ & $0.054$ & $0.005$ & B(A779) & $13.7$ & $3.98$ & $7.8$ \\
NGC 2892 & $-25.70$ & $11.77$ & $0.046$ & $237$ & $234$ & $0.013$ & $-0.154$ & $0.051$ & $0.004$ & I &  & $2.19$ & $2.2$ \\
NGC 3158 & $-26.28$ & $12.02$ & $0.255$ & $301$ & $289$ & $-0.034$ & $0.009$ & $0.032$ & $0.009$ & B & $13.3$ & $2.69$ & $9.5$ \\
NGC 3209 & $-25.55$ & $11.70$ & $0.039$ & $288$ & $247$ & $-0.146$ & ($-0.018)$ & $0.005$ & $-0.025$ & B & $11.8$ & $2.40$ & $2.7$ \\
NGC 3462 & $-25.62$ & $11.73$ & $0.085$ & $233$ & $214$ & $-0.048$ & $-0.326$ & $-0.017$ & $-0.015$ & I &  & $2.24$ & $2.5$ \\
NGC 3562 & $-25.65$ & $11.75$ & $0.038$ & $250$ & $241$ & $-0.069$ & $-0.140$ & $0.028$ & $-0.021$ & B & $13.5$ & $2.24$ & $8.3$ \\
NGC 3615 & $-25.58$ & $11.72$ & $0.399$ & $268$ & $232$ & $-0.232$ & $0.114$ & $-0.030$ & $-0.044$ & B & $13.6$ & $3.09$ & $5.1$ \\
NGC 3805 & $-25.69$ & $11.76$ & $0.496$ & $266$ & $225$ & $-0.262$ & $0.086$ & $0.019$ & $-0.060$ & S(A1367) & $14.8$ & $5.62$ & $430$ \\
NGC 3816 & $-25.40$ & $11.64$ & $0.105$ & $212$ & $191$ & $-0.016$ & $-0.142$ & $-0.038$ & $-0.055$ & S(A1367) & $14.8$ & $5.75$ & $1900$ \\
NGC 3842 & $-25.91$ & $11.86$ & $0.038$ & $262$ & $231$ & $-0.098$ & ($0.046)$ & $0.022$ & $-0.000$ & B(A1367) & $14.8$ & $5.89$ & $18000$ \\
NGC 3862 & $-25.50$ & $11.68$ & $0.045$ & $248$ & $228$ & $-0.069$ & $0.105$ & $-0.005$ & $-0.068$ & S(A1367) & $14.8$ & $5.89$ & $18000$ \\
NGC 3937 & $-25.62$ & $11.73$ & $0.074$ & $292$ & $243$ & $-0.150$ & ($0.042)$ & $0.015$ & $-0.002$ & B & $14.2$ & $5.89$ & $69$ \\
NGC 4073 & $-26.33$ & $12.05$ & $0.023$ & $316$ & $292$ & $-0.101$ & $0.199$ & $0.034$ & $0.043$ & B & $13.9$ & $4.37$ & $87$ \\
NGC 4472 & $-25.72$ & $11.78$ & $0.197$ & $292$ & $258$ & $-0.074$ & ($-0.088)$ & $0.023$ & $0.020$ & B(Virgo) & $14.7$ & $8.91$ & $1800$ \\
NGC 4555 & $-25.92$ & $11.86$ & $0.120$ & $328$ & $277$ & $-0.097$ & $-0.366$ & $0.044$ & $0.022$ & I &  & $5.89$ & $6.2$ \\
NGC 4816 & $-25.33$ & $11.61$ & $0.069$ & $217$ & $207$ & $-0.031$ & $-0.017$ & $0.003$ & $0.002$ & S(Coma) & $15.3$ & $13.18$ & $1900$ \\
\end{tabular}
\end{table*}
\begin{table*}
\setlength{\tabcolsep}{6pt}
\label{table:maincontinued}
\contcaption{}
\begin{tabular}{ll@{}rcccrrrrl@{\hspace{-1.3em}}rrr}
\hline
Galaxy & $M_K$ & $\log_{10} M_*$ & $\lambda_e$ & $\sigma_c$ & $\langle \sigma \rangle_e$ & $\gamma_{\rm inner}$ & $\gamma_{\rm outer}$ & $\langle h_4 \rangle$ & $h_4^\prime$ & env & $\log_{10} M_{\rm halo}$ & $1 + \delta_g$ & $\nu_{10}$ \\
 & [mag] & [$M_\odot$] &  & [km/s] & [km/s] &  &  &  &  &  & [$M_\odot$] &  & [$\overline{\nu}$] \\
(1) & (2) & (3) & (4) & (5) & (6) & (7) & (8) & (9) & (10) & (11) & (12) & (13) & (14) \\
\hline
NGC 4839 & $-25.85$ & $11.83$ & $0.053$ & $261$ & $275$ & $-0.026$ & ($0.298)$ & $0.061$ & $0.099$ & S(Coma) & $15.3$ & $13.18$ & $2600$ \\
NGC 4874 & $-26.18$ & $11.98$ & $0.070$ & $251$ & $258$ & $-0.030$ & $0.255$ & $0.046$ & $0.051$ & S(Coma) & $15.3$ & $13.18$ & $23000$ \\
NGC 4889 & $-26.64$ & $12.18$ & $0.032$ & $370$ & $337$ & $-0.128$ & $0.087$ & $0.051$ & $0.055$ & B(Coma) & $15.3$ & $13.18$ & $18000$ \\
NGC 4914 & $-25.72$ & $11.78$ & $0.054$ & $233$ & $225$ & $-0.023$ & $-0.213$ & $0.005$ & $-0.032$ & I &  & $1.12$ & $1.1$ \\
NGC 5129 & $-25.92$ & $11.86$ & $0.402$ & $260$ & $222$ & $-0.172$ & $0.203$ & $0.024$ & $0.029$ & I &  & $4.27$ & $4.8$ \\
NGC 5208 & $-25.61$ & $11.73$ & $0.615$ & $270$ & $235$ & $-0.226$ & ($0.467)$ & $0.001$ & $-0.024$ & B & $13.0$ & $5.01$ & $15$ \\
NGC 5322 & $-25.51$ & $11.68$ & $0.054$ & $246$ & $239$ & $-0.073$ & ($-0.178)$ & $-0.002$ & $-0.031$ & B & $13.7$ & $2.45$ & $20$ \\
NGC 5353 & $-25.45$ & $11.66$ & $0.551$ & $277$ & $225$ & $-0.141$ & ($0.277)$ & $0.018$ & $-0.037$ & B & $13.6$ & $2.63$ & $62$ \\
NGC 5490 & $-25.57$ & $11.71$ & $0.138$ & $349$ & $282$ & $-0.272$ & ($0.345)$ & $0.056$ & $0.020$ & I &  & $2.14$ & $9.5$ \\
NGC 5557 & $-25.46$ & $11.66$ & $0.035$ & $279$ & $223$ & $-0.153$ & ($0.120)$ & $0.015$ & $-0.056$ & B & $13.3$ & $2.57$ & $8.3$ \\
NGC 6223 & $-25.59$ & $11.72$ & $0.315$ & $274$ & $238$ & $-0.127$ & ($-0.304)$ & $0.008$ & $0.006$ & B & $13.5$ & $1.55$ & $6.0$ \\
NGC 6375 & $-25.53$ & $11.69$ & $0.240$ & $226$ & $187$ & $-0.128$ & ($-0.119)$ & $0.021$ & $-0.042$ & I &  & $1.17$ & $1.5$ \\
UGC 10918 & $-25.75$ & $11.79$ & $0.034$ & $247$ & $249$ & $-0.121$ & $0.185$ & $0.018$ & $0.051$ & I &  & $1.78$ & $4.7$ \\
NGC 6482 & $-25.60$ & $11.72$ & $0.137$ & $305$ & $291$ & $-0.066$ & $-0.174$ & $0.009$ & $-0.033$ & B & $13.1$ & $1.58$ & $1.0$ \\
NGC 6575 & $-25.58$ & $11.72$ & $0.124$ & $264$ & $234$ & $-0.098$ & ($-0.156)$ & $-0.011$ & $-0.000$ & I &  & $2.09$ & $4.9$ \\
NGC 7052 & $-25.67$ & $11.75$ & $0.148$ & $298$ & $266$ & $-0.093$ & $-0.199$ & $0.045$ & $0.017$ & I &  & $1.32$ & $0.8$ \\
NGC 7242 & $-26.34$ & $12.05$ & $0.037$ & $255$ & $283$ & $0.020$ & $0.118$ & $0.043$ & $0.031$ & B & $14.0$ & $6.31$ & $2700$ \\
NGC 7265 & $-25.93$ & $11.87$ & $0.039$ & $230$ & $206$ & $-0.112$ & $-0.034$ & $-0.003$ & $-0.061$ & B & $14.7$ & $6.92$ & $5100$ \\
NGC 7274 & $-25.39$ & $11.63$ & $0.090$ & $259$ & $244$ & $-0.089$ & $0.230$ & $0.030$ & $0.003$ & S & $14.7$ & $6.92$ & $3200$ \\
NGC 7386 & $-25.58$ & $11.72$ & $0.071$ & $312$ & $273$ & $-0.071$ & $-0.009$ & $0.031$ & $-0.003$ & B & $13.9$ & $2.57$ & $3.1$ \\
NGC 7426 & $-25.74$ & $11.79$ & $0.563$ & $284$ & $219$ & $-0.345$ & $0.099$ & $0.031$ & $-0.048$ & B & $13.8$ & $3.80$ & $8.3$ \\
NGC 7436 & $-26.16$ & $11.97$ & $0.085$ & $280$ & $263$ & $-0.092$ & $0.251$ & $0.043$ & $0.022$ & B & $14.4$ & $4.07$ & $100$ \\
NGC 7550 & $-25.43$ & $11.65$ & $0.038$ & $270$ & $224$ & $-0.117$ & $-0.373$ & $0.005$ & $-0.030$ & B & $11.9$ & $0.93$ & $1.0$ \\
NGC 7556 & $-25.83$ & $11.83$ & $0.049$ & $253$ & $243$ & $-0.013$ & $0.093$ & $0.029$ & $0.049$ & B & $14.0$ & $2.00$ & $17$ \\
NGC 7618 & $-25.44$ & $11.65$ & $0.247$ & $292$ & $265$ & $-0.127$ & $-0.140$ & $0.022$ & $-0.014$ & B & $13.7$ & $3.16$ & $240$ \\
NGC 7619 & $-25.65$ & $11.75$ & $0.119$ & $325$ & $277$ & $-0.155$ & ($0.073)$ & $0.020$ & $-0.002$ & B & $14.0$ & $1.55$ & $21$ \\
NGC 7626 & $-25.65$ & $11.75$ & $0.034$ & $269$ & $250$ & $-0.042$ & ($-0.416)$ & $0.045$ & $-0.019$ & S & $14.0$ & $1.55$ & $21$ \\
\end{tabular}

  Column notes:
  (1) Galaxy name, in order of increasing right ascension (not listed) for consistency with previous MASSIVE papers.
  (2) Extinction-corrected 2MASS total absolute $K$-band magnitude.
  (3) Stellar mass estimated from $M_K$ (eq. 2 of \citealt{maetal2014}, from \citealt{cappellari2013}).
  (4) Spin parameter $\lambda$ within $R_e$.
  (5) Central fiber dispersion.
  (6) Average luminosity-weighted dispersion within $R_e$.
  (7) Power law slope of $\sigma(R)$ at 2~kpc.
  (8) Power law slope of $\sigma(R)$ at 20~kpc. When $R_{\rm max} < 20$~kpc, the power law slope at $R_{\rm max}$ is listed instead, in parentheses. IC~0310 is removed from the list of outer fits entirely (see text).
  (9) Average luminosity-weighted $h_4$ within $R_e$, with typical formal errors $\sim 0.01$.
  (10) Gradient in $h_4$, defined as $\Delta h_4 / \Delta \log_{10} R$.
  (11) Group membership according to the 2MRS HDC catalogue: "B" for BGGs, "S" for satellites, "I" for isolated galaxies with fewer than 3 group members.
  Membership in Virgo, Coma, Perseus, or Abell clusters is listed.
  (12) Halo mass according to the HDC catalogue, or from updated literature sources (see text) for Virgo, Coma, and Perseus.
  (13) Large-scale galaxy overdensity from the 2M++ catalogue.
  (14) Local density in units of the mean $K$-band luminosity density $\overline{\nu} \sim 2.8 \times 10^8\; {\rm L}_\odot \; {\rm Mpc}^{-3}$.
  (**) Additional columns will be included in the electronic version of the table, available as online supplementary material, including $\sigma(R)$ fit parameters $\gamma_1$, $\gamma_2$, and $\sigma_0$ and uncertainties on $\gamin$ and $\gamout$.
\end{table*}

The MASSIVE survey targets a volume-limited sample of 116 early-type galaxies (ETGs) in the northern hemisphere and away from the galactic plane,\footnote{The total is 115 galaxies after we remove NGC 7681, which our IFS data revealed as a close pair of less-luminous bulges \citep{vealeetal2017}, each below the magnitude cut of our survey.}
with stellar masses $M_* > 4 \times 10^{11} M_\odot$ (estimated from $K$-band magnitudes $M_K < -25.3$ mag) and distances $D < 108$ Mpc.
The galaxies were selected from the Extended Source Catalogue \citep[XSC;][]{jarrettetal2000} of the Two Micron All Sky Survey \citep[2MASS;][]{skrutskieetal2006}.
The sample selection, methodology, and science goals of the survey were described in \citet{maetal2014} (Paper I of the MASSIVE survey). 
Uncertainties in galaxy parameters such as $D$, $M_K$, $M_*$, and $R_e$ were discussed in \citet{maetal2014} and \citet{vealeetal2017,vealeetal2017b}.

Thus far we have observed 90 MASSIVE galaxies with the Mitchell/VIRUS-P Integral Field Spectrograph (IFS) at the McDonald Observatory \citep{hilletal2008}.
This IFS covers a large 107\arcsec$\times$107\arcsec\ field of view with 246 evenly-spaced 4\arcsec-diameter fibres and a one-third filling factor, which we use to obtain contiguous coverage by oberving each galaxy with three dither positions.
The spectral range of the IFS spans 3650\AA\ to 5850\AA, covering the Ca H+K region, the G-band region, H$\beta$, the Mgb region, and many Fe absorption features.  The instrumental resolution varies by factors of about 20 per cent over this wavelength range but is typically around 4.5\AA\ full width at half-maximum.

The spectra from individual fibres in the central regions of our galaxies typically have a signal-to-noise ratio (S/N) 
exceeding 50, and we use these single-fibre spectra directly in the kinematic analysis. 
Beyond the central regions, we combine the fibres with lower S/N into radial and azimuthal bins, folding across the major axis and combining symmetrical bins such that each resulting co-added spectrum reaches at least S/N = 20. 
The line-of-sight velocity distribution (LOSVD) is parametrized as a Gauss-Hermite series up to order 6, and we obtain the best-fitting velocity $V$, dispersion $\sigma$, and higher order moments $h_3$, $h_4$, $h_5$, and $h_6$ using the penalized pixel-fitting (pPXF) method of \citet{cappellariemsellem2004}. 
Details of our kinematic analysis such as spectral continuum modeling, stellar template fitting, and error determination were described in \cite{vealeetal2017} (Paper V).

Our results for four measures of galaxy environments -- group membership, halo mass $\mh$, large-scale galaxy density $\delta_g$, and local galaxy density $\nu_{10}$ -- were presented in \cite{vealeetal2017b} (Paper VII) and listed in Table~1. 
Briefly, we take group membership and halo mass information from the HDC catalogue of \cite{crooketal2007,crooketal2008}, which is based on the 2MRS sample of \citet{huchraetal2005a,huchraetal2005b}.
Based on whether a galaxy is in a group with at least 3 members in the HDC catalogue, we assign each of our galaxies to be ``Isolated'', a ``Satellite'' galaxy in a group, or ``Brightest Group Galaxy'' (BGG).
Classification as a BGG is based entirely on the $K$-band luminosity, and is not necessarily equivalent to being the central galaxy \citep[e.g.][]{skibbaetal2011,hoshinoetal2015}, and also does not distinguish cases where two galaxies are nearly matched in luminosity.
We use $\mh$ measured by the projected mass estimator \citep{heisleretal1985} from the HDC catalogue; the 15 isolated galaxies in our sample therefore do not have a halo mass estimate.
 For the well-studied clusters of Virgo, Coma, and Perseus, we replace the $\mh$ taken from the HDC catalogue with values from the literature: $\mh = 5.5 \times 10^{14} M_\odot$ for Virgo \citep{durrelletal2014,ferrareseetal2012,schindleretal1999}, $\mh = 1.8 \times 10^{15} M_\odot$ for Coma \citep{kuboetal2007,falcoetal2014,rinesetal2003}, and $\mh = 6.7 \times 10^{14} M_\odot$ for Perseus.
The luminosity-weighted large-scale density contrast $\delta_g$ is calculated with a smoothing scale of a few Mpc in \cite{carricketal2015} based on the 2M++ redshift catalogue of \cite{lavauxhudson2011}.
We define the local density $\nu_{10}$ as the luminosity density within a sphere out to the $10^{\rm th}$ nearest neighbour of the galaxy, based on a parent catalogue including all galaxies with $M_K < -23.0$ from the 2MASS Redshift Survey \citep{huchraetal2012}.
With that luminosity cut, the parent catalogue is nearly (but not entirely) complete over our survey volume; see Section 3.4 and Appendix A of \cite{vealeetal2017b} for details.


\section{Radial Profiles of Stellar Velocity Dispersion}
\label{sec:profiles}

\subsection{Quantifying $\sigma(R)$}
\label{sec:profiles-fits}

 \begin{figure*}
\begin{center}
\includegraphics[page=1]{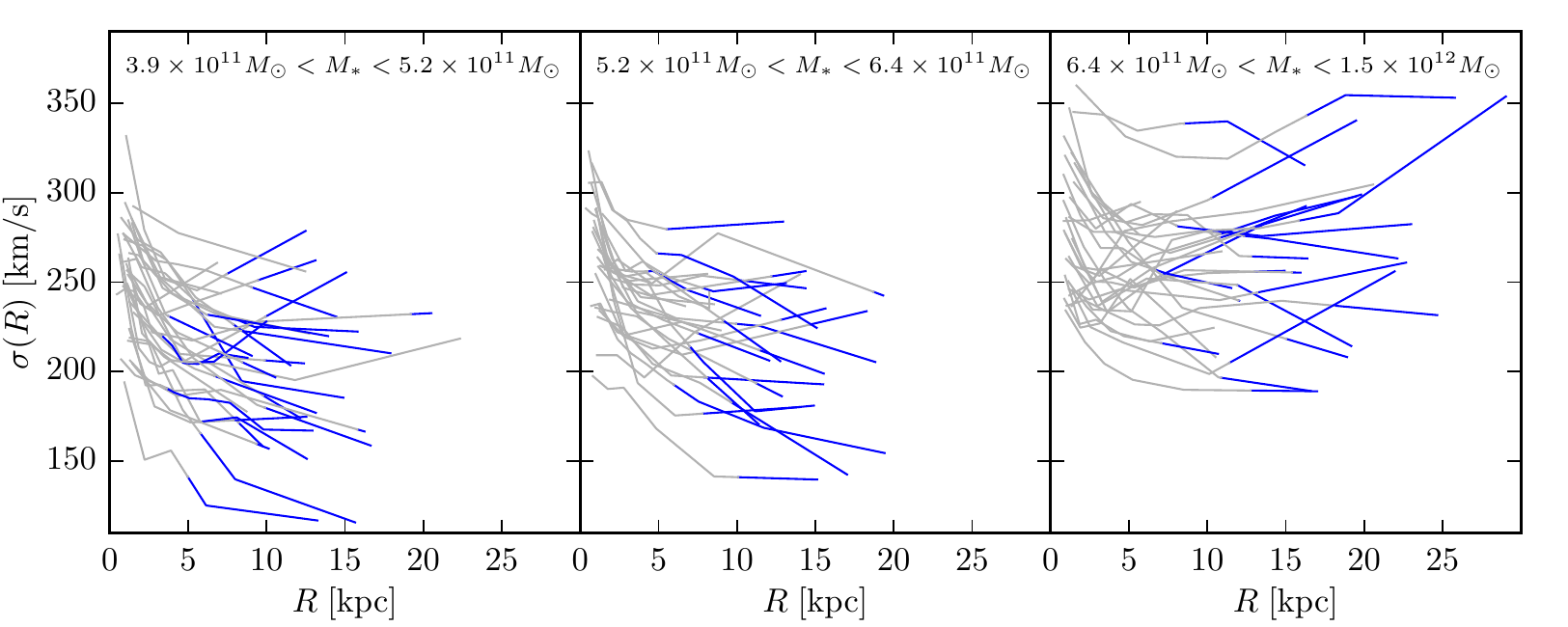}
\end{center}
\caption{Stellar velocity dispersion profiles of 90 MASSIVE galaxies, calculated as a luminosity-weighted average of individual data points in each annulus.
  The 3 panels are arranged from low to high $M_*$, in equal-number bins corresponding to $-25.31 {\rm ~mag} > M_K > -25.60 {\rm ~mag}$ (left), $-25.60 {\rm ~mag} > M_K > -25.82 {\rm ~mag}$ (center), and $-25.82 {\rm ~mag} > M_K > -26.64 {\rm ~mag}$ (right).
  Each profile is blue at $R>R_e$, and the outermost point represents the {\em average} radius of the outermost bin; the total radial extent of the data is up to $\sim 60\%$ larger.
  From left to right, the overall amplitude of $\sigma$ increases, with more profiles rising and fewer steeply falling.
}
\label{fig:profiles}
\end{figure*}

\begin{figure*}
\begin{center}
\includegraphics[page=1]{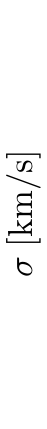}
\includegraphics[page=41]{plotting/figs-allprofiles.pdf}
\includegraphics[page=61]{plotting/figs-allprofiles.pdf}
\includegraphics[page=91]{plotting/figs-allprofiles.pdf}
\end{center}
\caption{
  Examples of velocity dispersion profiles, with individual data points (black and white points near positive and negative major axis respectively, grey points near minor axis) and profile fits (red).
  About 30\% of galaxies are fully consistent with a single power law fit (left panel), with 50\% having $\gamout > \gamin$ (middle panel) and 20\% having $\gamout < \gamin$.
    Most galaxies in the last category, by eye, appear to have a break radius smaller than the fiducial 5~kpc, and are often somewhat poor fits to the very outer data points (with the fit declining more steeply than the data).
    A compilation of this figure for all galaxies is shown in \aref{sec:appendix-allgals}.
  Galaxies where $R_{\rm max} < 20$~kpc have $\gamout$ measured at $R_{\rm max}$ (in parentheses) and are excluded from any statistics on $\gamout$.
}
\label{fig:examples}
\end{figure*}

The radial profiles of the stellar velocity dispersion, $\sigma(R)$, for the 90 MASSIVE galaxies are shown in \autoref{fig:profiles} in three increasing $M_*$ bins.
As $M_*$ increases (from left to right panels), the overall amplitude of $\sigma$ increases, and more $\sigma$ profiles become flat or rising in the outer parts instead of falling monotonically.

To quantify the overall shape of $\sigma(R)$, we fit a broken power law form as in \citet{vealeetal2017}:
\begin{equation}
  \label{eq:fit}
  \sigma(R) = \sigma_0 2^{\gamma_1 - \gamma_2} \left( \frac{R}{R_b} \right)^{\gamma_1} \left( 1 + \frac{R}{R_b} \right)^{\gamma_2 - \gamma_1}  \,,
\end{equation}
where the break radius $R_b$ is fixed at 5~kpc.
Although the apparent break radius varies somewhat below and above this value among different galaxies, degeneracies among $R_b$, $\gamma_1$, and $\gamma_2$ allow as good a fit for every galaxy with this fixed $R_b$ as with a free $R_b$.
Example fits are shown in \autoref{fig:examples}, and fits for all 90 galaxies are shown in \aref{sec:appendix-allgals}.

The parameters $\gamma_1$ and $\gamma_2$ in \autoref{eq:fit} represent the asymptotic logarithmic slopes of $\sigma(R)$, 
whose values can be more extreme
than the slopes of $\sigma(R)$ in the radial range covered by our IFS data.  
To account for this, we convert $\gamma_1$ and $\gamma_2$ to the {\it local} logarithmic slopes $\gamin$ and $\gamout$, where $\gamin$ is the power law slope of the fit at 2~kpc, and $\gamout$ is the power law slope of the fit at 20~kpc.
  We quote $\gamout$ only for the 56 galaxies ($\sim 2/3$ of the sample) where our data reach at least 20~kpc.\footnote{We also remove $\gamout$ for IC~0310, because it has unusually low $\sigma$ in the outer bins and is a significant outlier with $\gamout \sim -0.7$.}

Values of $\sigma$ (and error bars) come directly from the analysis of \citealt{vealeetal2017} (Paper V).
  The radial coordinates for each bin are a luminosity-weighted average over the radial coordinates of the fibre centres.
  For a single fiber near the centre of the galaxy, this can be quite inaccurate; a fibre of finite size with uniform flux that is exactly centred on the galaxy has an average radius of 2/3 the fibre radius, not 0.
  We adjust the radial coordinates of our data to account for this effect, assuming uniform flux across each fibre, and set the error on $R$ to 0.5\arcsec to account for astrometric uncertainty.
  These adjustments avoid potential problems with the fits due to steep power laws being very sensitive at small radius.

\autoref{fig:examples} illustrates three typical of $\sigma(R)$ shapes that we have found for MASSIVE galaxies: about 30\% are consistent with a single power law ($\gamin = \gamout$, left panel), 50\% have $\gamin < \gamout$ and typically fall to a minimum before rising at larger $R$ (middle panel), and 20\% have $\gamin > \gamout$ and are typically flat in the inner region then fall steeply at large $R$ (right panel).
Several galaxies in the last category are somewhat poor fits to the data in the very outskirts, with the fit declining more steeply than the data, but for the sake of uniformity we do not adjust the fits for these few galaxies.
Instead, we simply note that the most negative values of $\gamout$ in our sample belong to these galaxies, and are likely over-estimating the steepness of the decline at the edges of our data.

\begin{figure}
\begin{center}
\includegraphics[page=1]{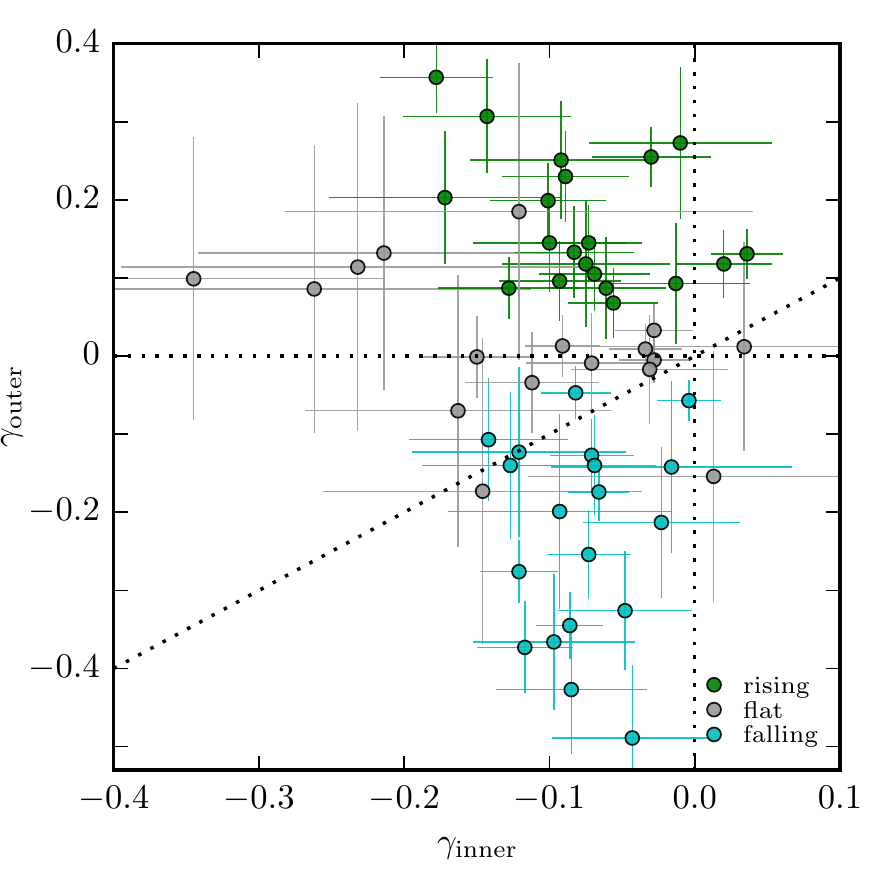}
\end{center}
\caption{Outer vs. inner logarithmic slopes of $\sigma(R)$ for the 56 galaxies with binned data extending to at least 20~kpc.
  Most $\gamin$ are negative, while $\gamout$ spans a range of both positive and negative values.
  We classify galaxies by their outer profile behavior: rising (green), falling (cyan), or flat (gray), where any galaxy consistent with $\gamout = 0$ (within errors) is considered flat.
  Galaxies with the most negative $\gamout$ tend to resemble NGC~1700 (right panel of \autoref{fig:examples}), where the fit likely over-estimates the steepness of the profile decline, but they are correctly identified as having a falling outer $\sigma(R)$ profile.
  Several galaxies are consistent with a single power law, $\gamin = \gamout$ (dotted line).
}
\label{fig:inout}
\end{figure}

\autoref{table:main} lists our measurements of $\gamin$ for all 90 MASSIVE galaxies and $\gamout$ for the 56 galaxies with a sufficiently large radial extent.
\autoref{fig:inout} shows $\gamin$ versus $\gamout$ for these 56 galaxies. 
We define rising, flat, and falling outer profiles to be those with $\gamout > 0$, $\gamout = 0$ within errors, and $\gamout < 0$ respectively.
The sample is split nearly equally among the three types, with 36\% rising, 30\% flat, and 34\% falling.

Uncertainties on $\gamin$ and $\gamout$ depend both on the uncertainty on the observed $\sigma$, and on whether the dispersion profile contains bumps, wiggles, or other features not accounted for in this simple fit.
Errors range from 0.01 to 0.2, with typical errors of around 0.05.
Errors for each galaxy are not tabulated in \autoref{table:main} but will be available in the electronic version of the table.
Flattened and/or fast rotating galaxies (e.g. NGC~5353) sometimes have a large spread in $\sigma$ at large radius, and localized bumps in $\sigma$ may be caused by the remains of merger activity \citep{schaueretal2014}.
  Although mergers for very massive galaxies are expected to be fairly common \citep[e.g.][]{edwardspatton2012,burkecollins2013}, galaxies with obvious interactions were removed in the original sample selection \citep{maetal2014} so we expect the overall impact of mergers to be relatively minor.
Some of the galaxies with $\gamin > \gamout$ might also be better fit by adding a second break radius, so that the center, intermediate radii, and outskirts could be fit by three distinct power laws.
Our main goal is to quantify the statistics of overall $\sigma(R)$ behavior in a uniform way across our sample, so we do not address these features further here.

\subsection{Effects of rotation: $\sigma$ versus $v_{\rm rms}$}
\label{sec:profiles-vrms}

\begin{figure}
\begin{center}
\includegraphics[page=1]{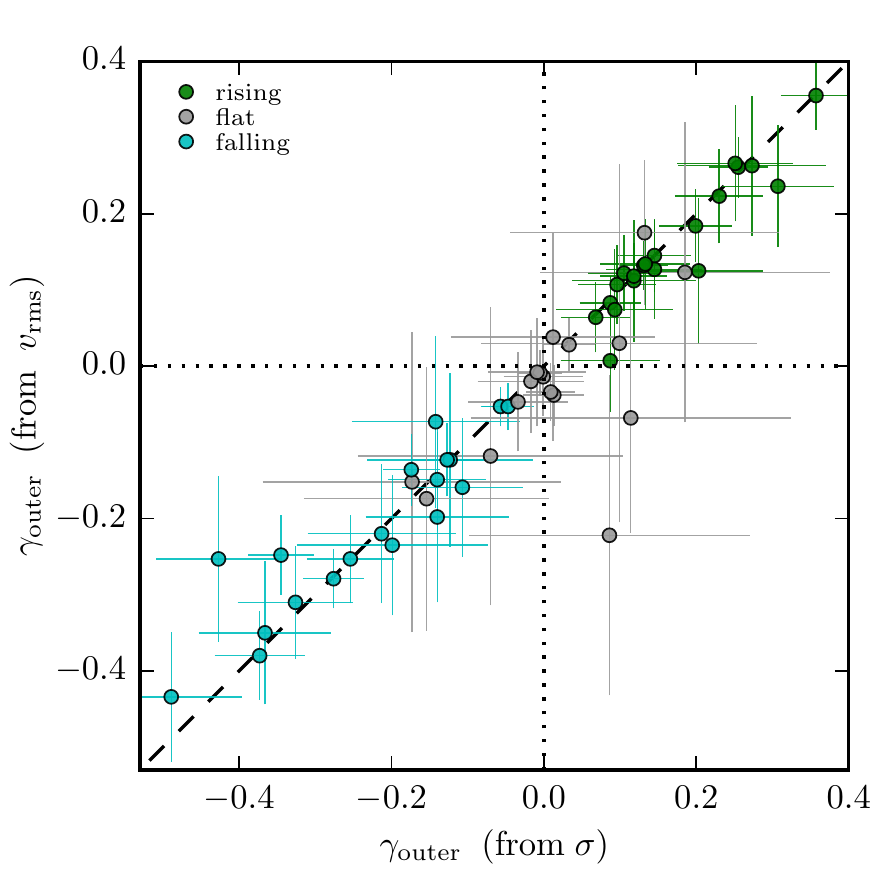}
\end{center}
\caption{
  Comparison of $\gamout$ for $v_{\rm rms}(R)$ and $\sigma(R)$.
  As expected, because $v_{\rm rms}$ is larger than $\sigma$ for outer bins of fast rotators, the outer profile of $v_{\rm rms}(R)$ is flatter than $\sigma(R)$ for a handful galaxies.
 However, the difference in $\gamout$ is small overall since most galaxies in our sample are slow or non-rotators.
}
\label{fig:vrms}
\end{figure}

Although many galaxies in the MASSIVE survey are slow or non-rotators \citep{vealeetal2017,vealeetal2017b}, 19 galaxies in the sample reported in this paper have spin parameter $\lambda_e \ge 0.2$ (see column 4 of \autoref{table:main}), where the galaxy rotation $V$ can contribute a non-negligible amount to the second velocity moment $v_{\rm rms} \equiv \sqrt{V^2 + \sigma^2}$.

To quantify the effects of rotation, we re-calculate $\gamout$ using $v_{\rm rms}(R)$ instead of $\sigma(R)$ and compare the two slopes in \autoref{fig:vrms}.
As expected, the effect of rotation is to make the slopes of the $v_{\rm rms}$ profile less extreme than those of the $\sigma$ profile.
The changes in $\gamout$ due to the difference in $v_{\rm rms}$ and $\sigma$, however, are generally small and affect less than 20\% of the sample.

Because the differences in slopes and profile fits between $v_{\rm rms}(R)$ and $\sigma(R)$ are generally small and do not impact our results presented below, we will use $\sigma(R)$ and the slopes computed from $\sigma(R)$ for the rest of the paper.

\subsection{The importance of a large field of view}
\label{sec:profiles-rmax}

\begin{figure}
\begin{center}
\includegraphics[page=2]{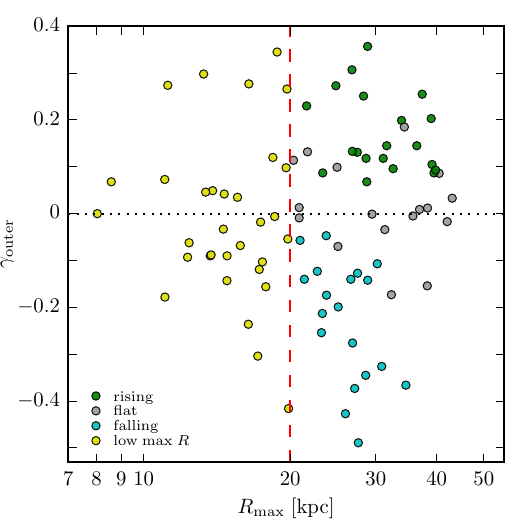}
\end{center}
\caption{
  Outer $\sigma(R)$ slope $\gamout$ versus $R_{\rm max}$, the maximum radius in kpc covered by our IFS data.
  The fraction of galaxies with rising outer profiles is smaller for galaxies with $R_{\rm max} < 20$~kpc, and many of those galaxies have uncertain fits (see \aref{sec:appendix-allgals}).
    For those galaxies, $\gamout$ is measured at $R_{\rm max}$ instead of at 20~kpc, so $\gamout$ cannot be fairly compared to the rest of the sample.
  About 2/3 of our galaxies have $R_{\rm max} > 20$~kpc, and we restrict our analysis to those galaxies to enable fair comparisons of $\gamout$ across the sample.
}
\label{fig:rmax}
\end{figure}

Our IFS data extend up to 40~kpc, with most galaxies in the 15-30~kpc range, corresponding to $\sim 1$ to $4R_e$.
This is usually far enough to capture the transition from falling inner $\sigma$ to rising outer $\sigma$ in \textsf{U}-shaped galaxies,
but a galaxy identified as \textsf{U}-shaped with data out to 30~kpc would likely appear to be monotonically decreasing if viewed to less than 10~kpc.  

\autoref{fig:rmax} shows how the outer profile behaviour $\gamout$ relates to the outer boundary of our IFS data $R_{\rm max}$.
Indeed, we find that galaxies with $R_{\rm max} < 20$~kpc are less likely to have rising outer profiles, and many of those with the smallest $R_{\rm max}$ have very few data points and uncertain fit quality.
They are included in the compilation of all 90 galaxies in \autoref{table:main} and \aref{sec:appendix-allgals}, with $\gamout$ in parentheses to indicate that it is measured at $R_{\rm max} < 20$~kpc rather than at 20~kpc.
In the following sections, we will use $\gamout$ only from the 56 galaxies with $R_{\rm max} > 20$~kpc.

Unusually small radial coverage of the binned data (in kpc) generally stems from some combination of distance (with closer galaxies having a smaller physical FOV) or poor observing conditions (making the signal-to-noise ratio too small in the galaxy outskirts), and so does not cause a bias in stellar mass or environment.
  About 1/3 of our galaxies are excluded from the analysis of $\gamout$ for having $R_{\rm max} < 20$~kpc, and we have verified that they reflect the overall sample in distribution of stellar mass and environment.

\subsection{Comparison with literature}
\label{sec:profiles-literature}

A number of the MASSIVE galaxies in this paper have published $\sigma(R)$ measurements, typically based on long-slit observations covering a smaller radial extent than our survey. 
In most cases, our results are reasonably consistent with these earlier measurements over comparable radial ranges, but our more extended data sometimes find  different slopes for $\sigma(R)$ in the previously unexplored outer regions.
In addition, our IFS data provide kinematic measurements over several angular bins at a given radius.
We discuss some specific cases here.

\citet{franxetal1989} found falling $\sigma(R)$ profiles for NGC~1700, NGC~4472, and NGC~7619.
We also find falling $\sigma(R)$ for all three galaxies, though with uniformly steeper power law slopes: $\gamin = -0.09$, $-0.07$, $-0.16$, instead of $-0.03$, $-0.02$, $-0.08$, respectively.
Our galaxies are measured to uniformly larger radii (29, 14, 11~kpc, instead of 10, 2, 9~kpc respectively), and each galaxy behaves differently at these larger radii.
NGC~1700 falls more steeply at large radius, while NGC~4472 keeps nearly the same power law slope and NGC~7619 becomes flat.

\citet{fisheretal1995} found falling $\sigma(R)$ profiles for NGC~2832, NGC~4073, NGC~4472, NGC~4874, NGC~4889, and NGC~7619, with NGC~4839 nearly flat.
Our inner slope $\gamin$ matches reasonably well with their power law slopes in all cases, where our (their) values are $-0.09$ $(-0.07)$, $-0.10$ $(-0.04)$, $-0.07$ $(-0.04)$, $-0.03$ $(-0.09)$, $-0.13$ $(-0.05)$, $-0.15$ $(-0.10)$, and $0.03$ $(-0.01)$, respectively.
We find a positive outer slope $\gamout$ for all of these except NGC~4472, and in every case our data extends to a larger radius (although not always to 20~kpc).
None of their measurements covered beyond 20~kpc except for NGC~2832, which went to $\sim 23$~kpc.
Our data for NGC~2832 extend to 33~kpc and find $\sigma$ to increase in this outer region.

Another series of papers presented long-slit data of some MASSIVE galaxies (typically to a radius of $\sim 20$\arcsec), but did not quote power-law slopes of the $\sigma$ profiles \citep{simienprugniel1997,simienprugniel1998,simienprugniel2000}:
NGC~0080, NGC~0410, NGC~0890, NGC~1573, NGC~1684, NGC~2340, NGC~3158, and NGC~5129 were included.
Many have large error bars on the $\sigma$ profiles, but when trends can be discerned, all galaxies have flat or falling $\sigma(R)$ and agree with our measurements where they overlap.  
The two possible exceptions are NGC~1684 and NGC~3158.
We see a large scatter in $\sigma$ in the outskirts of NGC~1684, but no coherent increase, so the few outer points from \citet{simienprugniel2000} are likely part of that scatter.
For NGC~3158, we do see a small bump in $\sigma$ at the outskirts of the \citet{simienprugniel1998} measurement, but we find that $\sigma$ remains flat beyond that radius.
Of these 8 galaxies, we find two galaxies (NGC~0080 and NGC~5129) to show rising $\sigma$ beyond 20\arcsec.

A few individual galaxies have also been studied previously: NGC~1600 \citep{verolmeetal2002}, NGC~2320 \citep{crettonetal2000}, and NGC~2672 \citep{bonfantietal1995}, and they similarly agree with our measurements where they overlap.
None of those galaxies show rising $\sigma$ in our sample.
On the other hand, of the five galaxies with data in both this paper and \citet{loubseretal2008} (NGC~2832, NGC~3842, NGC~4839, NGC~4874, and NGC~4889), only NGC~3842 does not show a rising outer profile in our data.
NGC~4839 shows a rising $\sigma(R)$ but with limited radial bins, and the remaining three all have clear rising outer $\sigma(R)$.
In all three cases, our data extend farther than that in \citet{loubseretal2008} and agree with their data in the inner regions.

Many galaxies in the Coma cluster have falling $\sigma(R)$ \citep{thomasetal2007}, but that paper found all three Coma galaxies in the MASSIVE sample (NGC~4839, NGC~4874, and NGC~4889) to have at least some hint of a rising profile.
Of those, NGC~4839 had the most prominent rise, and was also the only galaxy to be measured to (and slightly beyond) $R_e$. 
In our sample, the situation is reversed, with NGC~4839 measured to less than 10~kpc while NGC~4874 and NGC~4889 are measured to at least 30~kpc and show a substantial rise in $\sigma$.

We have also compared those of our galaxies that overlap with ATLAS$^{\rm 3D}$ (NGC~4472, NGC~5322, and NGC~5557) in Appendix B of \citealt{vealeetal2017} and find that all the kinematics (not just $\sigma$) match well where they overlap in radius.
None of these three galaxies are in the subset of ATLAS$^{\rm 3D}$ galaxies studied in the SLUGGS survey \citep{fosteretal2016}.


\begin{figure*}
\begin{center}
\includegraphics[page=1]{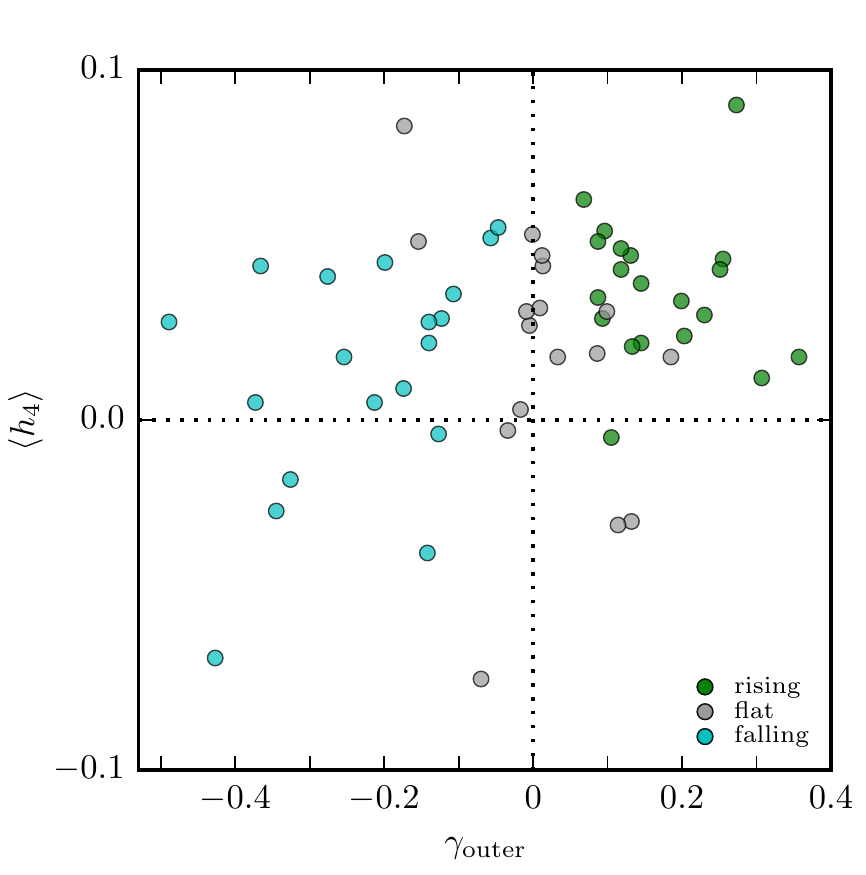}
\includegraphics[page=2]{plotting/figs-h4.pdf}
\end{center}
\caption{Luminosity-weighted average $h_4$ within $R_e$ (left) and $h_4$ gradient (right) versus $\gamout$.
  Nearly all galaxies with rising outer $\sigma(R)$ profiles have positive $\langle h_4 \rangle$.
There is a positive correlation between $h_4$ gradient and $\gamout$ (dashed line in right panel) with a slope of $0.096 \pm 0.037$, and the corresponding $p$-value for the significance of the correlation is $p = 0.012$.
The few outliers in both panels are perhaps due to limitations on spectral resolution, as they are galaxies with low overall $\sigma$.
}
\label{fig:h4}
\end{figure*}

\section{Dispersion profiles versus velocity kurtosis $h_4$}
\label{sec:kinematics}

When interpreting dispersion profiles, the degeneracy between velocity anisotropy and enclosed mass (or mass-to-light ratio) can be somewhat alleviated by examining the kurtosis $h_4$ of the LOSVD.
Radial anisotropy is generally associated with positive $h_4$ and lower projected $\sigma$, while tangential anisotropy is associated with negative $h_4$ and higher projected $\sigma$ \citep{gerhardetal1998}.
Even in isotropic systems, however, positive $h_4$ can also arise from gradients (of either sign) in circular velocity \citep{gerhard1993,baesetal2005}.
A transition from the galaxy dispersion to cluster dispersion can also be interpreted as a galaxy surrounded by intracluster light (ICL) from a diffuse halo of stars controlled by the cluster gravity, although a clear decomposition into these two components is difficult \citep{benderetal2015}.
In practice, for cluster BCGs (especially cD galaxies) an ICL component and a circular velocity gradient play similar roles for our purposes, both producing a positive $h_4$.

\autoref{fig:h4} shows that, as we found in \citet{vealeetal2017}, MASSIVE galaxies have generally positive $h_4$ and there is evidence for a correlation between outer $\sigma$ gradient and $h_4$ gradient.
The $p$-value for the significance of the correlation (with slope $0.096 \pm 0.037$) is $0.012$.
One feature not found in \citet{vealeetal2017} are the few galaxies with negative $\langle h_4 \rangle$ and substantially negative ($< -0.1$) gradients in $h_4$.
These are likely due to low $\sigma$, with $\sigma < 200$~\kms causing large scatter in $h_4$ because of limitations on wavelength resolution.
In particular, the most negative two points in both panels of \autoref{fig:h4} and the outlying positive $h_4$ gradient are all galaxies with low average $\sigma$.
Because \citet{vealeetal2017} focused only on the most massive 41 galaxies of the sample, it is not surprising that outliers related to low $\sigma$ did not arise in that subsample.
Formal errors on $\langle h_4 \rangle$ are typically $\sim 0.01$, but additional systematic effects are likely present for low $\sigma$.

As in \citet{vealeetal2017}, we argue the correlation between the $h_4$ gradient and $\gamout$ is much more likely to be a consequence of circular velocity gradients (or an ICL component) than velocity anisotropy.
If all galaxies had similar mass (and light) profiles, then a positive $h_4$ gradient related to greater radial anisotropy at large radius would be expected to accompany a more {\em negative} $\sigma$ gradient, and so cannot explain the observed correlation.
It is also true that tangential anisotropy can only boost $\sigma$ by a limited amount, and in particular cannot boost it above the circular velocity.
This makes invoking radial anisotropy to explain low $\sigma$ more easily justified in most cases than invoking tangential anisotropy to explain high $\sigma$.
For these reasons, we argue that variations in the total mass profiles across our sample (and variations from isothermal profiles) are likely present.

In this context, our positive $\langle h_4 \rangle$ could result from either mass profiles ($V_{\rm circ}$ gradients, ICL components) or radial anisotropy, or some combination of both.
If mass profiles are the dominant effect on $h_4$, then in principle {\em tangential} anisotropy may also be common in our sample in spite of the overall positive $h_4$.
Based on our data, we can make no claims about the anisotropy of our sample as a whole, or about changes in anisotropy across our sample, without more detailed dynamical modeling.
We leave such modeling for future papers.


\begin{figure*}
\begin{center}
\includegraphics[page=1,width=0.99\columnwidth]{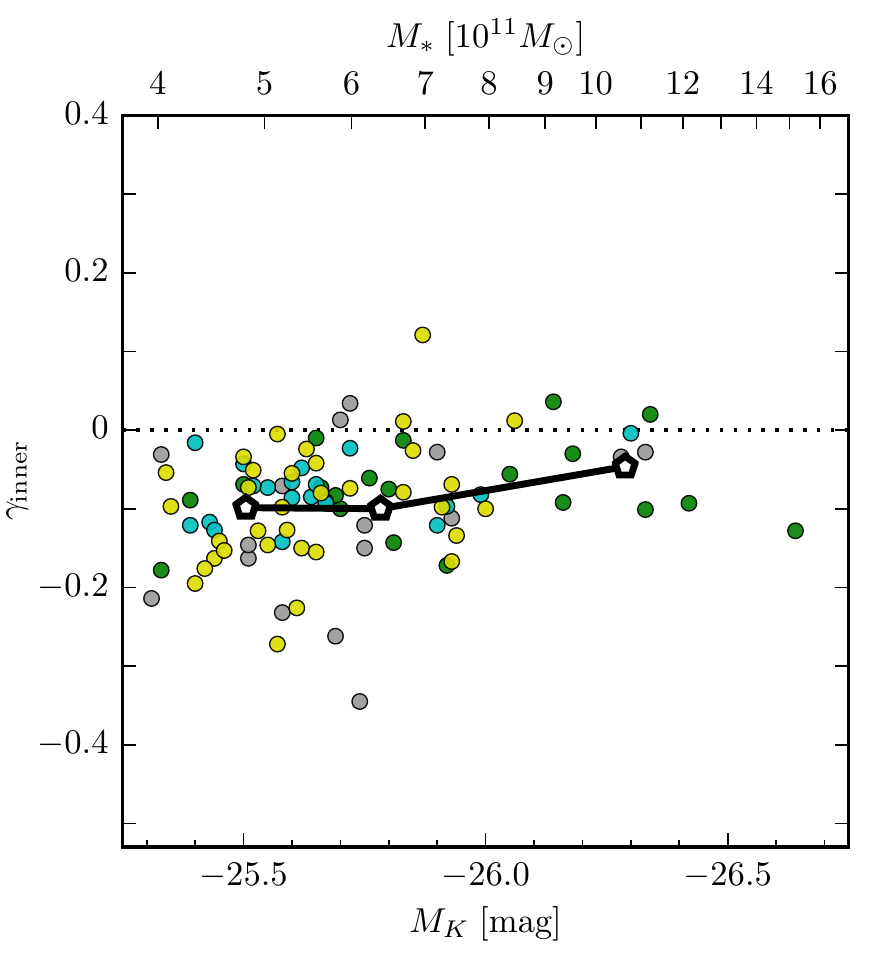}
\includegraphics[page=2,width=0.99\columnwidth]{plotting/figs-mk.pdf}
\end{center}
\caption{Inner (left) and outer (right) slopes of $\sigma(R)$ versus $M_*$ (and $M_K$).
  The inner $\sigma$ profiles are mostly falling, in the range of $-0.2 < \gamin < 0$, with an average power-law slope (thick black pentagons) of -0.1 becoming more shallow at -0.05 in the highest $M_*$ bin.
  Galaxies with $R_{\rm max} < 20$~kpc (yellow; left out of $\gamout$ sample) have a similar distribution of $\gamin$ as the other galaxies.
The outer slopes $\gamout$ span a larger range of about $-0.4$ to $+0.3$ at the low end of the mass range,
but are mostly positive at $M_* \ga 8\times 10^{11} M_\odot$.  
Rising outer profiles are found across the whole mass range, whereas falling outer profiles are found only for galaxies with $M_* \la 8\times 10^{11} M_\odot$, leading to a correlation between average $\gamout$ and $M_*$. 
}
\label{fig:mstargammas}
\end{figure*}

\section{Dispersion profiles versus galaxy mass and environment}
\label{sec:environment}

\subsection{$\sigma$ profiles and stellar mass $M_*$}
\label{sec:environment-mstar}

\begin{figure}
\begin{center}
\includegraphics[page=3]{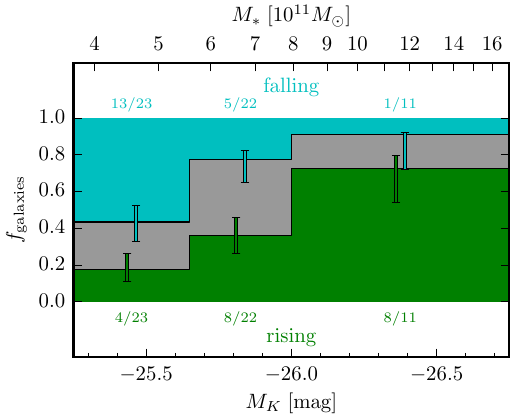}
\end{center}
\caption{Fraction of rising (green),
  flat (grey), and falling (cyan) outer $\sigma$ profiles versus $M_K$ and $M_*$.
  In spite of the narrow mass range of our sample, there is a strong correlation between outer $\sigma(R)$ profile behaviour and stellar mass.
}
\label{fig:mstarhist}
\end{figure}

\autoref{fig:mstargammas} shows the power law slope of $\sigma(R)$ at 2~kpc ($\gamin$, left) and 20~kpc ($\gamout$, right) versus stellar mass $M_*$ (and $M_K$) for the 90 MASSIVE galaxies (56 of which have data reaching to 20~kpc; see \autoref{sec:profiles-rmax}).
Most galaxies have falling $\sigma(R)$ in the inner part
($\gamin\la 0$), but the outer $\sigma(R)$ has slopes spanning $-0.4 \la \gamout \la 0.3$.
Thick black pentagons indicate the average values of $\gamin$ and $\gamout$ for each of three $M_*$ bins.
Both $\gamin$ and $\gamout$ show a positive trend with $M_*$, but the trend with $\gamout$ is much steeper.
  In particular, only one of the 11 most massive galaxies has a falling profile, and it is very shallow.
  A direct linear fit gives a slope of $0.056 \pm 0.030 \; {\rm mag}^{-1}$ for $\gamin$ versus $M_K$, and $0.170 \pm 0.083 \; {\rm mag}^{-1}$ for $\gamout$ versus $M_K$.
  A simple linear model is clearly inappropriate for the relationship with $\gamout$ versus $M_K$, where the overall scatter relative to the fit is $\sim 0.19$ (much larger than the typical errors on $\gamout$) and varies substantially with $M_K$.
  To investigate the relationship another way, \autoref{fig:mstarhist} shows the fraction of galaxies with rising (green), flat (grey), and falling (cyan) outer $\sigma(R)$ for the three $M_*$ bins.
The error bars\footnote{Error bars are calculated as described in Appendix~B of \citet{vealeetal2017b}, with a Beta distribution as prior and posterior.
We use $n_{\rm prior} = 2$ for a very weak prior, and $\mu_{\rm prior}$ is the total sample fraction of rising (20/56) or falling (19/56) profiles.}
in \autoref{fig:mstarhist} reflect the uncertainty due to small number statistics in each bin.

As discussed in \autoref{sec:kinematics}, detailed mass modeling would be needed to determine whether the correlations between $\gamout$ and $M_*$ seen in \autoref{fig:mstargammas} and \autoref{fig:mstarhist} indicate mass profile changes, or velocity anisotropy variations, or both.
Dynamical and lensing measurements have found roughly isothermal total mass profiles for many elliptical galaxies, especially within one or two $R_e$ \citep{gerhardetal2001,treuetal2006,koopmansetal2009,augeretal2009,augeretal2010,thomasetal2011,sonnenfeldetal2013,cappellarietal2015}.
But there is some evidence that smaller galaxies may have steeper than isothermal mass profiles \citep{romanowskyetal2003,napolitanoetal2009,deasonetal2012,morgantietal2013,alabietal2016}, whereas more massive galaxies may have shallower than isothermal profiles \citep{newmanetal2013}.
Simulations have suggested that merger histories may influence velocity anisotropy \citep{dekeletal2005,wuetal2014}, so the fact that more massive galaxies have a more extensive merger history may also link mass to velocity anisotropy.

In all of these cases, it is important to keep in mind the physical scale under consideration: at small radii, central supermassive black holes become important, while stars dominate at intermediate radii and dark matter dominates at large radii.
For example, core scouring by merging supermassive black holes likely impacts the $\sigma$ profiles near the galaxy center \citep{thomasetal2014} and may relate to $\gamin$.
The IFS data presented here do not resolve the galaxy cores, however, so we will not consider further the central regions of the galaxy and $\gamin$.
In the following sections we focus instead on how the outer $\sigma$ profiles (i.e. $\gamout$) may correlate with galaxy environment and dark matter.

\subsection{Outer $\sigma$ profiles vs. environmental measures $\mh$, $\delta_g$, and $\nu_{10}$}
\label{sec:environment-mhdeltanu}

\begin{figure*}
\begin{center}
\includegraphics[page=1]{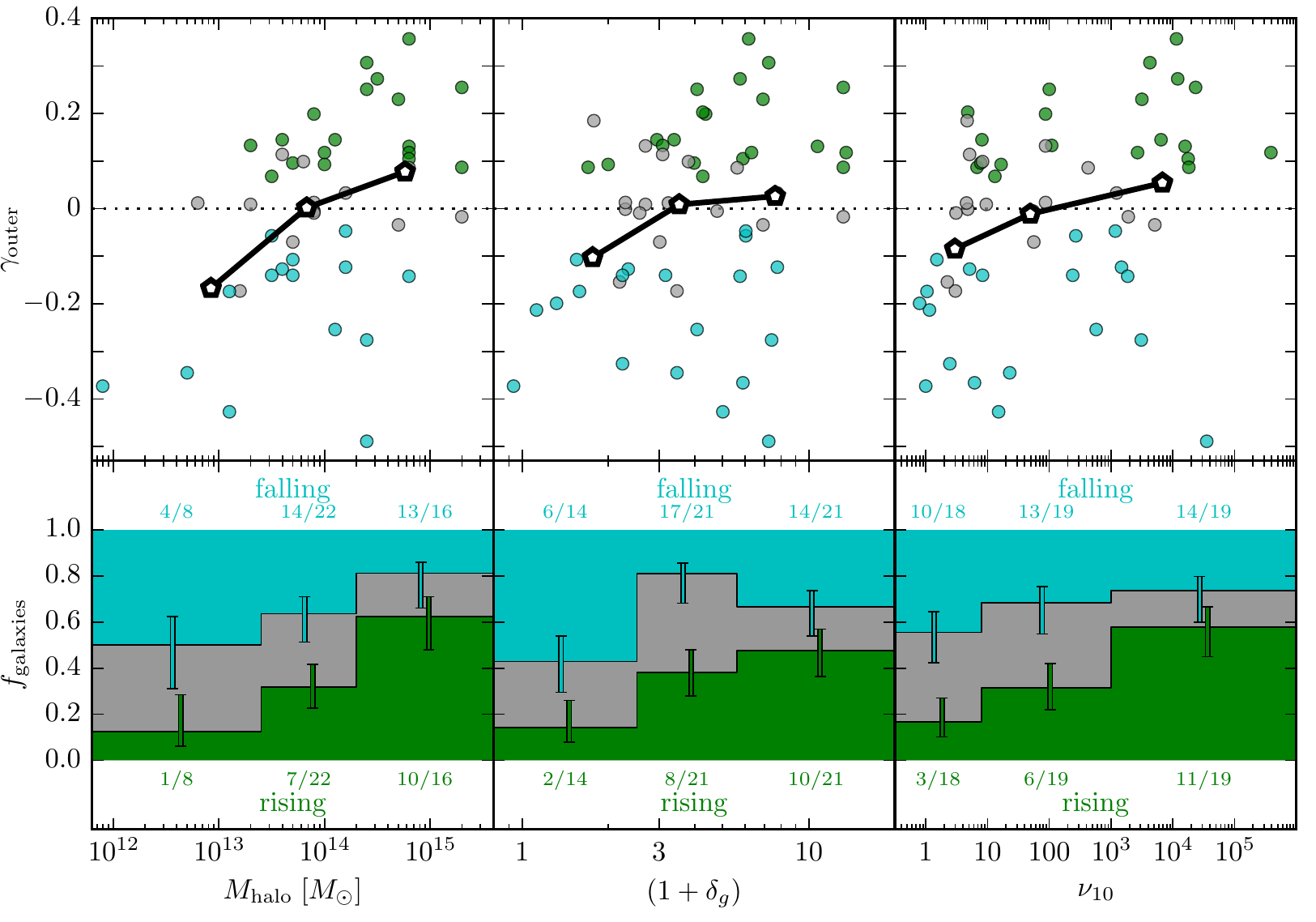}
\end{center}
\caption{
Outer $\sigma(R)$ profile behavior versus three environment measures: $\mh$ (left), large-scale density $\delta_g$ (middle), and local density $\nu_{10}$ (right).
The top panels show $\gamout$; the bottom panels show the fraction of rising (green), flat (grey), and falling (cyan) outer profiles.
For all three environmental measures, $\frise$ and average $\gamout$ (open black pentagons) increase in more dense environments, and $\ffall$ decreases correspondingly. 
  The most steeply rising outer profiles are also found in the highest density environments, while steeply falling profiles are found in a wide range of environments.
}
\label{fig:mhdeltanu}
\end{figure*}

\begin{table}
  \caption{Two-sample Kolmogorov-Smirnov test $D$ statistic and $p$ values for comparing the distribution in $M_*$ or environment of the galaxy sample, split into two samples of galaxies by separating first rising profiles (left columns) and then falling profiles (right columns).}
  \label{table:ks}
  \begin{center}
\begin{tabular}{l|cc|cc}
 & \multicolumn{2}{c}{(rising v. flat+falling)} & \multicolumn{2}{c}{(rising+flat v. falling)} \\
 & $D$-statistic & $p$-value & $D$-statistic & $p$-value \\
\hline
$M_K$ & 0.428 & 0.012 & 0.414 & 0.019 \\
$M_{\rm halo}$ & 0.385 & 0.032 & 0.338 & 0.088 \\
$\delta_g$ & 0.361 & 0.052 & 0.263 & 0.301 \\
$\nu_{10}$ & 0.467 & 0.005 & 0.289 & 0.205 \\
\end{tabular}
\end{center}

\end{table}

\autoref{fig:mhdeltanu} shows the power law $\sigma$ slope at 20~kpc ($\gamout$, upper panels) and the fraction of galaxies with rising, flat, and falling outer $\sigma$ profiles (lower panels) versus three environmental variables: halo mass $\mh$ (left), large-scale overdensity $\delta_g$ (middle), and local density $\nu_{10}$ (right). 
Within each upper panel, the average $\gamout$ for three environmental bins is shown as thick black pentagons.
All panels show clear trends of increasing $\frise$, increasing average $\gamout$, and decreasing $\ffall$ in denser environments, and the trend is most prominent for $\mh$.
Simple linear fits of $\gamout$ versus each environment variable give distinctly positive slopes ($0.126 \pm 0.036$, $0.193 \pm 0.093$, and $ 0.043 \pm 0.017 \; {\rm dex}^{-1}$ for $\mh$, $\delta_g$, and $\nu_{\rm 10}$ respectively), similar to the case with $M_K$ in the previous section, and again the scatter around each linear fit is much larger than the typical errors on $\gamout$.

The upper panels of \autoref{fig:mhdeltanu} also show that galaxies with steeply rising outer $\sigma(R)$ are found preferentially in high density environments, while galaxies with steeply declining outer $\sigma(R)$ span a wide range of environments.
In terms of halo mass, the galaxies with rising $\gamout$ have an average $\mh$ of $4.7 \times 10^{14} M_\odot$, about four times
more massive than the average $\mh$ of $1.2 \times 10^{14} M_\odot$ for galaxies with falling $\gamout$.

We have performed a two-sample Kolmogorov-Smirnov (KS) test to compare the distributions of galaxy mass and environments, first comparing galaxies with rising profiles to those with flat or falling profiles, and second comparing galaxies with falling profiles to those with flat or rising ones.
\autoref{table:ks} shows both the $D$ statistic, which measures the maximum difference between the empirical cumulative distribution functions of the two samples, and the $p$-value, which quantifies the significance of that difference taking into account sample size.
\footnote{We have adjusted the $p$-value of the KS test for distributions in $\mh$ to account for the reduced sample size, due to some galaxies being isolated with no measurement of $\mh$ available, by using the full sample size including isolated galaxies when converting the $D$-statistic to a $p$-value.
This reduces the $p$-values from $\sim 0.06, 0.17$ to $\sim 0.03, 0.09$.}

We find that the most significant difference between the samples is related to the increase in $\frise$ with increasing $\nu_{10}$, with $p \sim 0.005$.
The next two most significant differences are related to stellar mass, with $\frise$ increasing ($p \sim 0.012$) and $\ffall$ decreasing ($p \sim 0.019$) at higher mass/brighter $M_K$.
There is an interesting feature of separating strongly rising and falling profiles from flat ones: while $M_*$ has comparable $p$-values whether separating rising profiles or falling ones, the environment measures tend to have much worse $p$-values for separating falling profiles.
This indicates that different mechanisms may be behind the trends, and we will discuss this further in \autoref{sec:conclusions}.

Much like the situation with $M_*$ in \autoref{sec:environment-mstar}, the correlation between $\sigma$ profile behaviour and environment may be related to changing mass and luminosity profiles, changing velocity anisotropy, or both.
Based on our results in \autoref{sec:kinematics}, we find it unlikely that anisotropy alone can be behind the correlation.
Instead, we interpret this as evidence that galaxies in more massive haloes may have total mass profiles that are shallower than isothermal, while those in less massive haloes have the more ``typical'' isothermal profiles.
Some lensing results have also suggested this \citep{newmanetal2015}.
We cannot entirely rule out anisotropy playing some role, however.
The connection between merging history and anisotropy may also result in a connection between environment and anisotropy, since galaxies in more dense environments likely experience a more extensive merger history.
This parallels our discussion from \autoref{sec:environment-mstar}, and indeed the correlation between $M_*$ and environment makes the two arguments equivalent to some degree.
We will explore in the next sections how to distinguish whether trends with environment are simply a reflection of trends in $M_*$, or vice versa, or whether both are independent.

\subsection{Outer $\sigma$ profiles vs. group membership}
\label{sec:environment-bgg}

\begin{figure*}
\begin{center}
  \includegraphics[page=2]{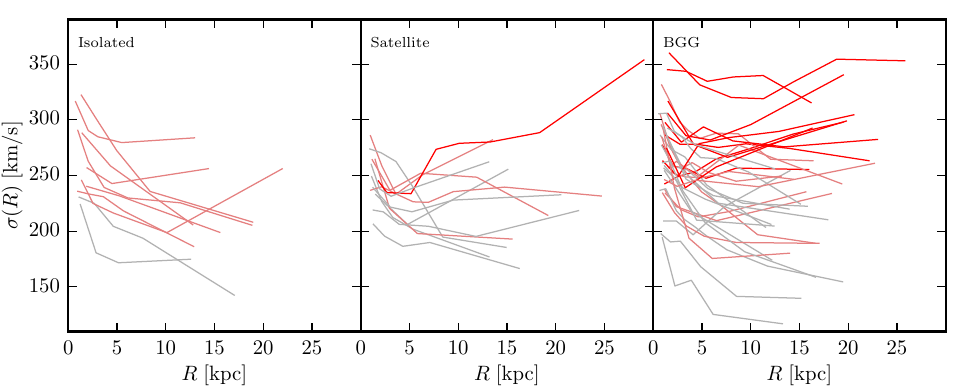}
\end{center}
\caption{
  Dispersion profiles as in \autoref{fig:profiles}, but separated by group membership status.
  The line colors correspond to the 3 bins of $M_*$ in \autoref{fig:mstarhist}, with grey for the least massive and red for the most massive.
 The outermost point represents the {\em average} radius of the outermost bin; the total radial extent of the data is up to $\sim 60\%$ larger.
Ten of the most massive 11 galaxies in our sample are BGGs;
the exception is NGC~4874, the second brightest galaxy in the Coma cluster.
}
\label{fig:envprofiles}
\end{figure*}

\begin{figure*}
\begin{center}
  \includegraphics[page=1]{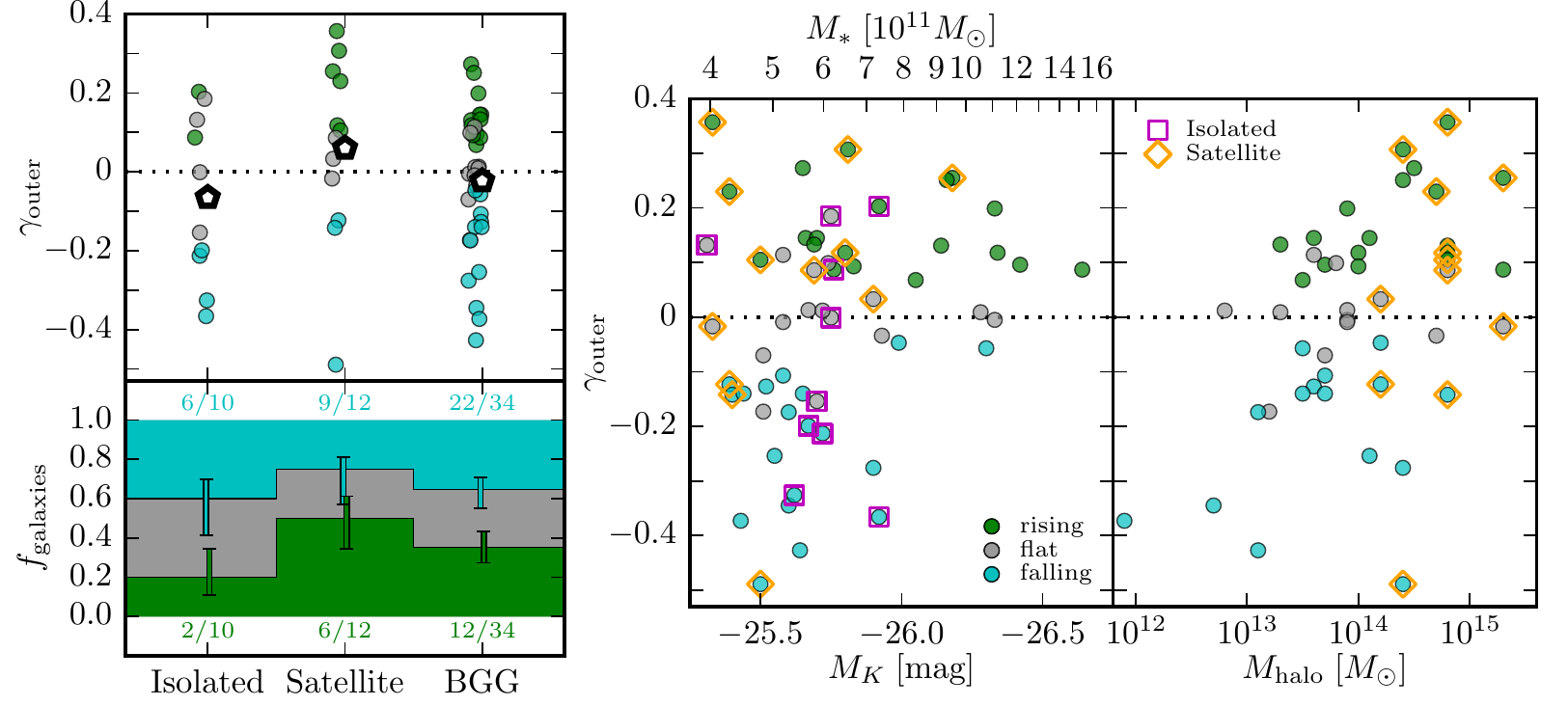}
\end{center}
\caption{
Outer slope $\gamout$ and fraction of rising (green), flat (grey), and falling (cyan) galaxies
versus group membership status (left panels), and $\gamout$ versus stellar and halo masses with group membership highlighted (right panels).
The average $\gamout$ (open black pentagons) is slightly higher for satellite galaxies (yellow diamonds in right panel), which include the most steeply rising $\sigma$ profiles.
  }
\label{fig:bgg}
\end{figure*}

\autoref{fig:envprofiles} and \autoref{fig:bgg} show $\sigma(R)$ for the 90 MASSIVE galaxies sorted by group membership, i.e., whether it is the brightest galaxy or a satellite in a group/cluster, or it is relatively isolated.
The differences in the fractions of rising or falling profiles for the three types are small in comparison to the differences with $M_*$ or $\mh$.

The right side of \autoref{fig:bgg} shows the $\gamout$ versus $M_*$ and $\mh$ panels of \autoref{fig:mstargammas} and \autoref{fig:mhdeltanu}, but with group membership marked with additional symbols (magenta squares are isolated; orange diamonds are satellites; unmarked galaxies are all BGGs).
The satellite galaxies in our sample, as expected, are found only in the higher mass haloes ($\mh \ga 10^{14} M_\odot$); in the lower mass haloes, only the BGGs are luminous enough to pass our survey magnitude cutoff.
Satellite galaxies also reside in the lower part of the $M_*$ range.
The average $\gamout$ for satellite galaxies is slightly higher than that for BGGs or isolated galaxies, reflecting the typical $\gamout$ of galaxies in massive haloes more than the typical behaviour of low mass galaxies.
Many of the most steeply rising profiles are found in satellite galaxies.

The physical significance of rising $\sigma$ in BGG and satellite galaxies may be different, especially if rising $\sigma$ is connected to the total mass profiles.
In many discussions (e.g. \citealt{benderetal2015}) the focus is on a central galaxy in a large cluster, and the total mass profile is simply the combination of stellar mass and the dark matter of the cluster halo.
Ongoing detailed mass modeling of our satellite galaxies
will help elucidate the impact of the group or cluster halo mass on a non-central galaxy, which may have its own dark matter subhalo or be moving at a substantial velocity relative to the group or cluster rest frame.
Some of our BGGs may also not be central galaxies (e.g. \citealt{skibbaetal2011,hoshinoetal2015}), and cases where the two brightest galaxies are nearly matched in luminosity are also of interest.

\autoref{fig:bgg} shows that 2 of the 10 isolated galaxies in our sample with $\gamout$ measured at 20~kpc have distinctly rising profiles: NGC~2693 and NGC~5129.
NGC~2693 is only marginally inconsistent with $\gamout=0$, with $\gamout = 0.087 \pm 0.065$, while NGC~5129 has $\gamout = 0.203 \pm 0.085$.
Both galaxies are at fairly large distances (74 and 108 Mpc, respectively).
If we assume they are BGGs of groups whose rank 3 members fall just below the 2MASS limit, they will have luminosity gaps ($\sim 2.0$~mag and $2.7$~mag respectively) comparable to those found for some other BGGs in our sample.
Only about 10 of our 85 galaxies have luminosity gaps (between rank 1 and rank 3) as large as 2.0~mag, and only 5 of those gaps as large as 2.5~mag.
All 10 of those galaxies have $\mh < 10^{14} M_\odot$.
These two isolated galaxies would thus be unusually fossil-like if they reside in halos of $\mh \ga 10^{14} M_\odot$.
If they instead reside in smaller halos, their group composition may be comparable to that of many BGGs in our sample, but they would then have $\gamout$ as high or higher than all our other BGGs at those halo masses.

\subsection{Untangling the joint relationships of $M_*$ and $\mh$}
\label{sec:environment-test}

\begin{figure*}
  \begin{center}
    \includegraphics[page=1]{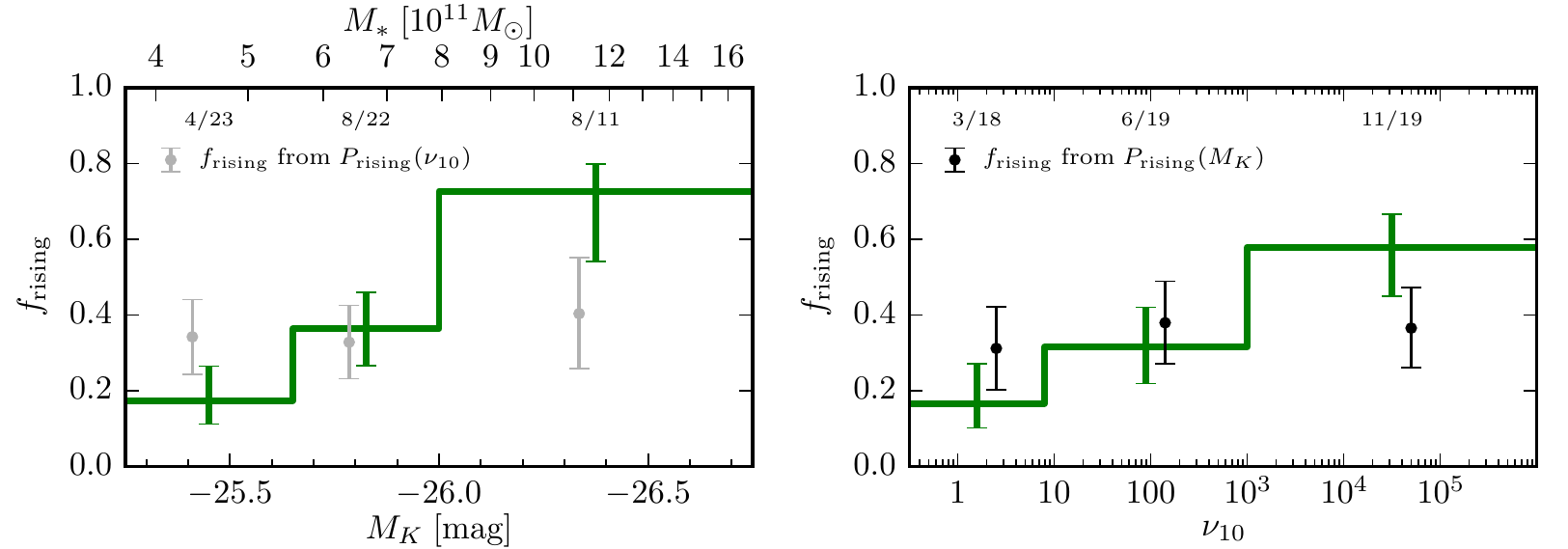}
\end{center}
  \caption{
  Comparing the assumptions that the probability of having a rising $\sigma(R)$ profile ($P_{\rm rising}$) is a function of $\nu_{10}$ only (gray; left panel), or a function of $M_K$ only (black; right panel).
    In each case $P_{\rm rising}$ is a logistic function fit to the data, with each galaxy at 0 (falling) or 1 (rising), so it is independent of the choice of binning.
    $P_{\rm rising}$ is used to calculate $\frise$ via 1000 Monte Carlo trials (see text).
    This is compared to the actual measured $\frise$ (green lines), with points offset slightly in the $x$-direction for clarity.
   The assumption that $P_{\rm rising}$ is a function only of $\nu_{10}$ does not predict the observed sharp increase in $\frise$ with $M_K$ (left panel). 
    Likewise, the assumption that $P_{\rm rising}$ is a function only of $M_K$ produces only a tiny increase in $\frise$ versus $\mh$, not enough to match the data (right panel).  Both stellar mass and environment therefore drive the observed trend in $\frise$.
}
\label{fig:test}
\end{figure*}

\begin{figure*}
  \begin{center}
    \includegraphics[page=2]{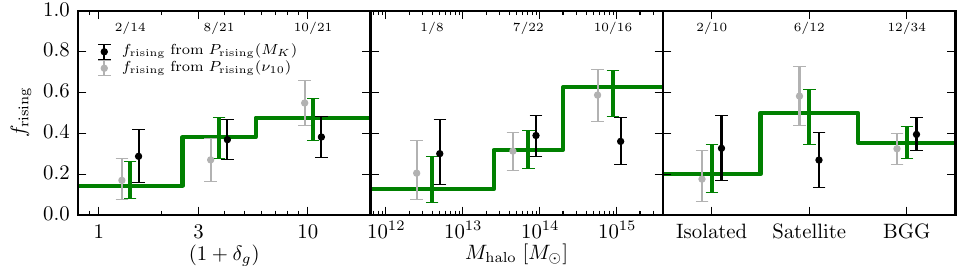}
\end{center}
  \caption{
    Comparing how well $\frise$ versus $\delta_g$, $\mh$ and group membership status is reproduced under the assumption that $P_{\rm rising}$ is a function of $M_K$ only (black) or of $\nu_{10}$ only (gray).
    The gray points can reproduce the actual measured $\frise$ in every case, as might be expected due to the correlations among different measures of environment.
    Points within each bin are offset slightly for clarity.
}
\label{fig:envtest}
\end{figure*}

In this section we consider two opposing, extreme assumptions: first, that the probability of a galaxy having a rising $\sigma$ profile (denoted $P_{\rm rising}$) is a function of $M_*$ (or $M_K$) only, and second, that it is a function of $\nu_{10}$ only.
If $P_{\rm rising} = P_{\rm rising} (M_*)$, and $\nu_{10}$ plays no direct/independent role in influencing the $\sigma$ profile, we still expect some residual correlation between the measured fraction of rising profiles ($\frise$) and $\nu_{10}$.
This arises because the distribution of $M_*$ changes with $\nu_{10}$ for our sample, so convolving $P_{\rm rising}(M_*)$ with that distribution gives a fraction of rising profiles that depends on $\nu_{10}$.
The reverse also applies: if $P_{\rm rising} = P_{\rm rising} (\nu_{10})$ then the connection between $\nu_{10}$ and $M_*$ results in some correlation between $\frise$ and $M_*$.

\autoref{fig:test} compares the ``predictions'' of each of these extreme assumptions with the actual measured $\frise$.
The test samples used to calculate $\frise$ for the two assumptions (i.e. using $P_{\rm rising}(M_*)$ and $P_{\rm rising}(\nu_{10})$) are constructed by assigning each galaxy in our sample a {\em probability} of having a rising profile, then running 1000 Monte Carlo trials assigning falling/rising profiles according to those probabilities and counting the resulting fractions in each bin.
$P_{\rm rising} (M_*)$ is constructed by fitting a logistic function to the unbinned $M_*$ data (where falling profiles are 0 and rising profiles are 1), and similarly for $\nu_{10}$.

In each case, the test sample accounts for only a tiny amount of the increase in $\frise$ with $M_*$ or $\nu_{10}$,
indicating that both stellar mass and environment are responsible for the trends in $\frise$.
The relatively small number statistics in each bin result in substantial estimates of the error, so the actual $\frise$ and test sample $\frise$ in each individual bin have errors overlapping at least slightly.
  However, the overall behaviour of $\frise$ versus $M_*$ and $\nu_{10}$ is very different between the nearly flat test samples and the correlations in the actual data.
This is in contrast to the case of fast and slow rotators in \citet{vealeetal2017b}, where $M_*$ alone determines $P_{\rm slow}$ (within errors).
It also differs from the case of $\ffall$ in this paper; unsurprisingly, since the change in $\ffall$ is much steeper for $M_K$ than $\nu_{10}$, we find that using $P_{\rm falling} (M_*)$ to construct $\ffall (\nu_{10})$ is consistent with the observed fraction, while using $P_{\rm falling} (\nu_{10})$ to construct $\ffall (M_*)$ does not account for the observed fraction.

We can also ask whether $P_{\rm rising} (M_*)$ or $P_{\rm rising} (\nu_{10})$ does better at ``predicting'' the fraction of galaxies with rising profiles as a function of the other environment measures.
\autoref{fig:envtest} shows that $P_{\rm rising} (\nu_{10})$ generates $\frise$ versus $\delta_g$, $\mh$, and group membership that matches better to the data.
  This is not surprising because $\nu_{10}$ is very well correlated with $\mh$ and thus with group membership status, and also correlates with $\delta_g$ very closely at low density.


\section{Conclusions}
\label{sec:conclusions}

We have measured and characterized the line-of-sight stellar velocity dispersion profiles $\sigma(R)$ of 90 early-type galaxies in the MASSIVE survey, spanning $K$-band magnitude $-25.3 > M_K > -26.7$, or stellar mass $4 \times 10^{11} M_\odot < M_* < 2 \times 10^{12} M_\odot$.  
Our IFS data cover 107\arcsec$\times$107\arcsec\ field of view and extend up to 40 kpc in radii, with most galaxies in the 15-30 kpc range.
We find a radial coverage to $\sim 20$~kpc or beyond is necessary to measure the outer dispersion profiles of local massive ETGs in a consistent way across the sample (\autoref{fig:rmax}).

We find a variety of shapes for $\sigma(R)$: monotonically falling and rising profiles, \textsf{U}-shaped profiles that fall to a minimum before flattening or rising at large radius, and profiles that are nearly flat in the center and fall at large radius.
We quantify the shapes of $\sigma(R)$ with an inner logarithmic slope at 2~kpc ($\gamin$) and an outer logarithmic slope at 20~kpc ($\gamout$).
Of the 90 galaxies in our sample, 56 have binned data reaching at least 20~kpc, and we limit our analysis of $\gamout$ to those galaxies.
All but 1 galaxy in our sample have $\gamin$ that is negative or consistent with zero (\autoref{fig:inout} and \autoref{table:main}), and most are in the range $-0.2 < \gamin < 0$.
By contrast, $\gamout$ ranges from $\sim -0.5$ to $+0.4$, where 36\% have rising outer $\sigma(R)$ ($\gamout > 0$), 30\% have flat profiles ($\gamout = 0$ within errors), and 34\% have falling profiles ($\gamout < 0$; \autoref{fig:inout}).

We show that the fraction of galaxies with rising outer $\sigma(R)$ profiles increases significantly over our $M_*$ range (\autoref{fig:mstargammas}, \autoref{fig:mstarhist}).  
That fraction also increases in denser environments, as quantified by halo mass $\mh$, large-scale density $\delta_g$, and local density $\nu_{10}$ (\autoref{fig:mhdeltanu}).
Among those, the trend is most prominent with $\mh$ and $\nu_{10}$, with the steepest rising profiles belonging to galaxies in the most dense environments.

We find that the probability of a galaxy having a rising outer $\sigma$ profile cannot be adequately expressed as either a function of $M_*$ alone or as a function of $\nu_{10}$ alone (\autoref{fig:test}).
Both $M_*$ and environment are therefore responsible for driving the trends in the rising fraction.
This is to be contrasted with galaxy rotation in the previous MASSIVE paper \citep{vealeetal2017b}, where the sharp increase in the fraction of slow rotators with $M_*$ is enough to explain the correlation between galaxy rotation and all environmental measures.
The first and second velocity moments of massive galaxies therefore reflect different aspects of their past assembly histories.

We find a higher fraction of galaxies with $\gamout > 0$ among satellite galaxies than among BGGs, and a lower fraction among isolated galaxies (\autoref{fig:bgg}).
  Rising dispersion profiles are not limited to BGGs but are found for all three types, although the two isolated galaxies with rising profiles may be the BGGs of groups only slightly more ``fossil-like'' than others in our sample (\autoref{sec:environment-bgg}).
Compared to the BGGs, the satellite galaxies in our sample 
occupy the lower part of the $M_*$ range (middle panel of \autoref{fig:bgg}) but the upper part of the $\mh$ range (right panel of \autoref{fig:bgg}).
Since the fraction of rising dispersion profiles increases with both $M_*$ and $\mh$ for the galaxies in our study as a whole, these two factors have a competing influence on the fraction of satellites with rising profiles.
The satellite galaxies appear to follow the typical behaviour of galaxies at high $\mh$ rather than low $M_*$, but our sample size is not large enough to say difinitively whether satellite galaxies truly have fewer $\sigma(R)$ profiles with $\gamout < 0$ at fixed mass than the overall sample.
Future larger surveys can provide important new insights by comparing the kinematics of BGGs, satellites, and isolated galaxies at {\it fixed} $M_*$ and $\mh$.

We find a positive correlation between the outer $\sigma$ gradient $\gamout$ and the gradient of the LOSVD kurtosis $\Delta h_4 / \Delta \log R$ (\autoref{fig:h4}).
Based in part on this correlation, we argue that the rising $\sigma$ profiles seen in our galaxies and the trends with $M_*$ and environment are likely caused at least in part by variations in total mass profiles (including variations from isothermal).
It is unlikely that rising $\sigma$ profiles can be explained by tangential velocity anisotropy alone, but a positive gradient in circular velocity is consistent with both positive $\sigma$ gradients and positive $h_4$ gradients.
Our results can accommodate a range of velocity anisotropy, so long as any tangential anisotropy (associated with negative $h_4$) is not extreme enough to overcome gradients in circular velocity (associated with positive $h_4$) as the primary influence on $\langle h_4 \rangle$, which we find to be generally positive.

Most likely, both mass profile shape and velocity anisotropy play a role in determining the $\sigma$ profile.
More detailed modeling is required to make any definitive statements, but the correlations with galaxy mass and environment suggest at least one possible scenario.
\autoref{fig:mhdeltanu} shows an apparently sharp cutoff of the maximum allowed $\gamout$ that increases with $\mh$, and to a lesser extent with $\delta_g$ and $\nu_{10}$.
Perhaps this is because the galaxy environment controls the underlying total mass profile shape (in central galaxies) and the presence of non-equilibrium motions (in satellites moving with respect to the cluster), with the surrounding dark matter dictating the maximum possible rise of outer $\sigma (R)$.
\autoref{fig:mstargammas} shows a similar cutoff for the {\em minimum} allowed $\gamout$, which increases with $M_K$.
Perhaps this is because substantial radial anisotropy, likely needed to explain very steeply falling $\sigma$, cannot survive the extensive merger histories typical of very massive galaxies.
At the most extreme masses nearly every galaxy may converge on a homologous anisotropy profile - not necessarily isotropic, but with less extreme radial anisotropy at large radii - while lower mass galaxies show a range of anisotropy profiles.
This scenario is consistent with the fact that our observed $h_4$ implies some variation in mass profiles and can accommodate a range of anisotropy.
It is also consistent with how the classification of ``flat'' profiles impacts the correlations, with $\frise$ correlating more strongly than $\ffall$ with $\mh$, while $\ffall$ correlates more strongly than $\frise$ with $M_*$.

This scenario is again speculative, based on qualitative arguments about the connections among enclosed mass, $\sigma$, velocity anisotropy, and $h_4$ that have been noted in the literature.
Our ongoing dynamical modeling efforts using the reported kinematics will provide deeper insights
into the trends reported in this paper.


\section*{Acknowledgements}

We thank Marijn Franx for useful discussions, and the anonymous referee for helpful comments.
The MASSIVE survey is supported in part by NSF AST-1411945, NSF AST-1411642, HST-GO-14210, and HST-AR-1457.




\bibliographystyle{mnras}
\bibliography{massive_viii}




\appendix

\section{Individual profiles}
\label{sec:appendix-allgals}

Each galaxy in our sample is fit to \autoref{eq:fit} as described in \autoref{sec:profiles-fits}.
\autoref{fig:allgals1} shows the $\sigma$ profiles for all 90 galaxies (in the same order as \autoref{table:main}), with the best-fit profile overlaid in red.

\begin{figure*}
\begin{center}
\includegraphics[page=1]{plotting/figs-allprofiles.pdf}
\includegraphics[page=2]{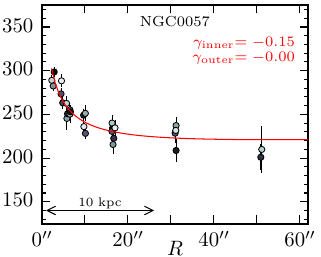}
\includegraphics[page=3]{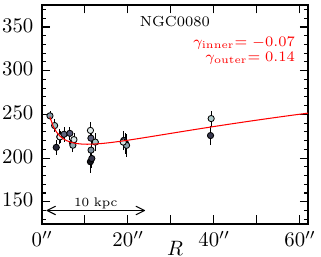}
\includegraphics[page=4]{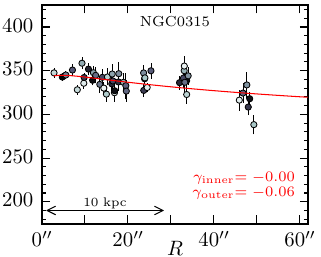}
\\
\includegraphics[page=1]{plotting/figs-allprofiles.pdf}
\includegraphics[page=5]{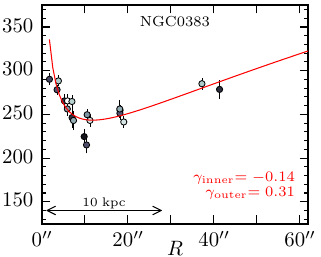}
\includegraphics[page=6]{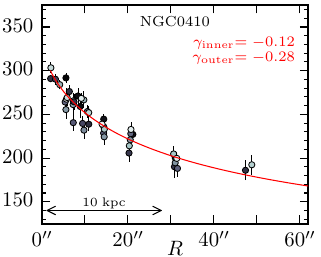}
\includegraphics[page=7]{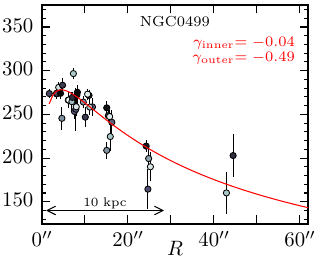}
\\
\includegraphics[page=1]{plotting/figs-allprofiles.pdf}
\includegraphics[page=8]{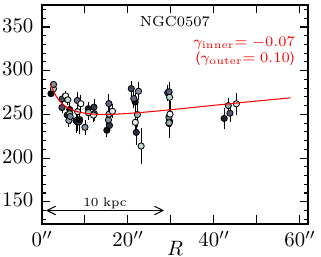}
\includegraphics[page=9]{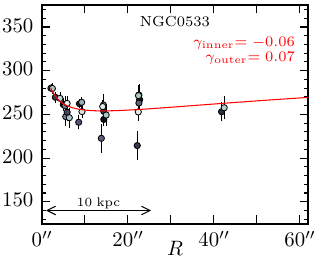}
\includegraphics[page=10]{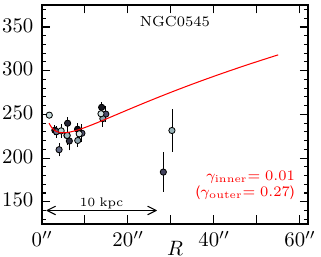}
\\
\includegraphics[page=1]{plotting/figs-allprofiles.pdf}
\includegraphics[page=11]{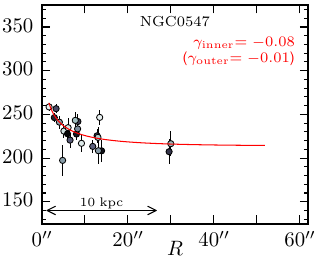}
\includegraphics[page=12]{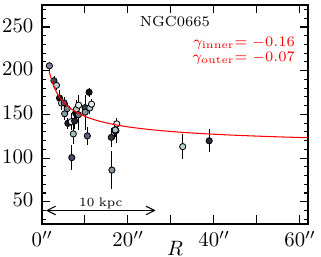}
\includegraphics[page=13]{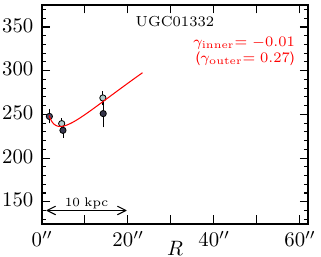}
\\
\includegraphics[page=1]{plotting/figs-allprofiles.pdf}
\includegraphics[page=14]{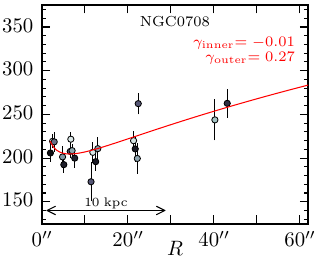}
\includegraphics[page=15]{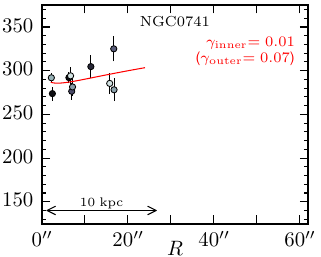}
\includegraphics[page=16]{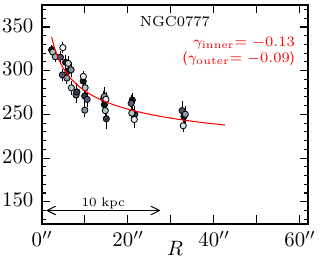}
\end{center}
\caption{Individual galaxy dispersion profiles. Within each galaxy, $\sigma$ for each spatial bin is shown with error bars, with point color corresponding to angle from major axis; black and white correspond to positive and negative major axis, and gray corresponds to the minor axis. Double power law fits are shown in red, with the local power law slope at 2~kpc ($\gamma_{\rm inner}$) and 20~kpc ($\gamma_{\rm outer}$) from \autoref{table:main}. If the maximum extent of the galaxy data is less than 20~kpc, the local slope at $R_{\rm max}$ is shown in parentheses.}
\label{fig:allgals1}
\end{figure*}
\begin{figure*}
\begin{center}
\ContinuedFloat
\includegraphics[page=1]{plotting/figs-allprofiles.pdf}
\includegraphics[page=17]{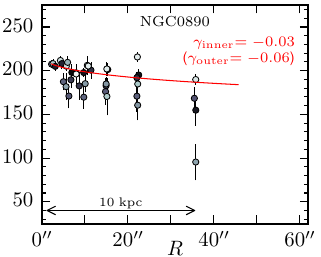}
\includegraphics[page=18]{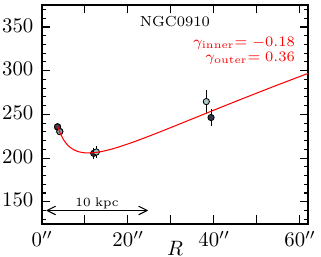}
\includegraphics[page=19]{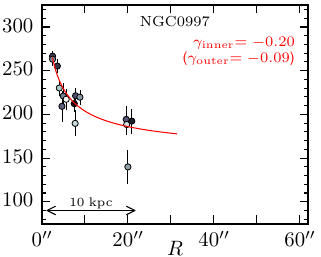}
\\
\includegraphics[page=1]{plotting/figs-allprofiles.pdf}
\includegraphics[page=20]{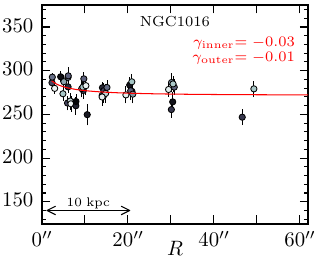}
\includegraphics[page=21]{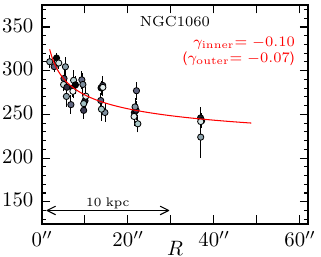}
\includegraphics[page=22]{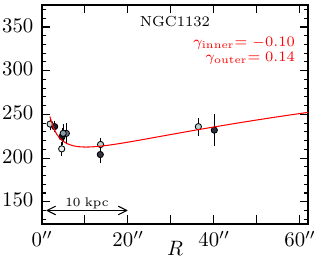}
\\
\includegraphics[page=1]{plotting/figs-allprofiles.pdf}
\includegraphics[page=23]{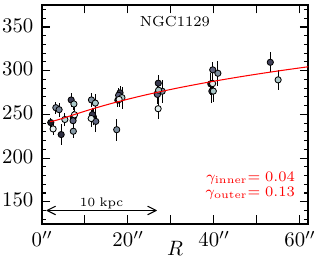}
\includegraphics[page=24]{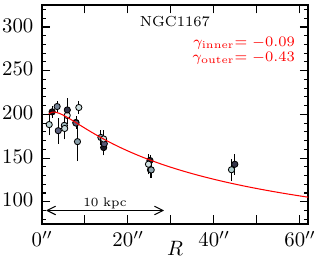}
\includegraphics[page=25]{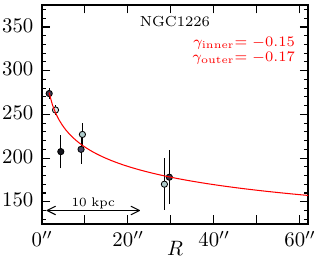}
\\
\includegraphics[page=1]{plotting/figs-allprofiles.pdf}
\includegraphics[page=26]{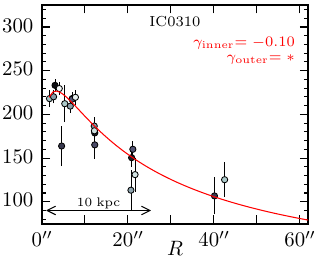}
\includegraphics[page=27]{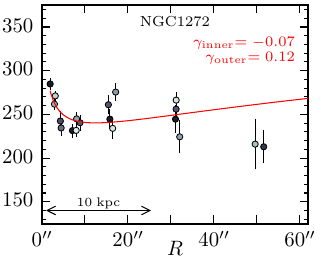}
\includegraphics[page=28]{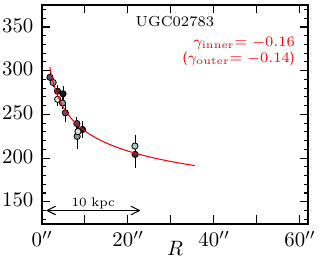}
\\
\includegraphics[page=1]{plotting/figs-allprofiles.pdf}
\includegraphics[page=29]{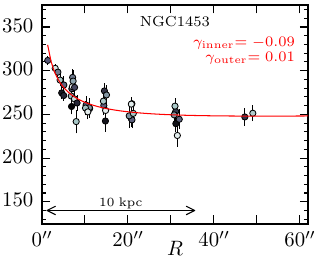}
\includegraphics[page=30]{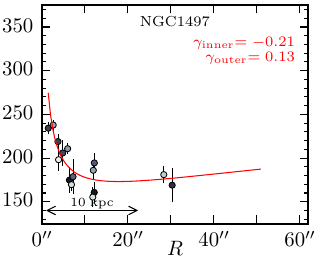}
\includegraphics[page=31]{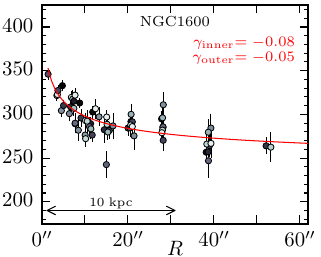}
\end{center}
\caption{(continued)}
\label{fig:allgals2}
\end{figure*}
\begin{figure*}
\begin{center}
\ContinuedFloat
\includegraphics[page=1]{plotting/figs-allprofiles.pdf}
\includegraphics[page=32]{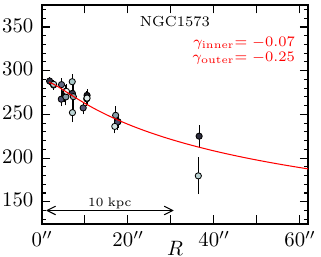}
\includegraphics[page=33]{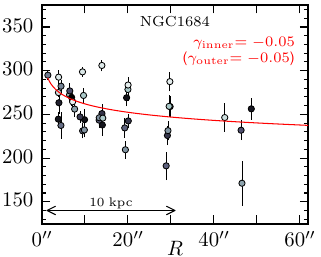}
\includegraphics[page=34]{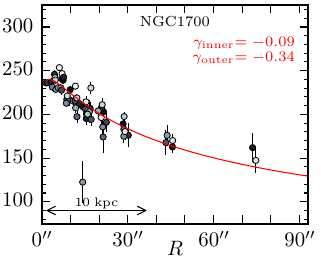}
\\
\includegraphics[page=1]{plotting/figs-allprofiles.pdf}
\includegraphics[page=35]{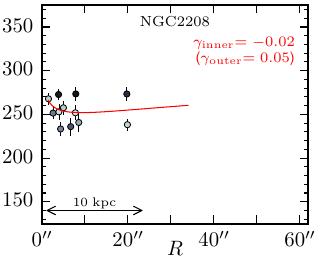}
\includegraphics[page=36]{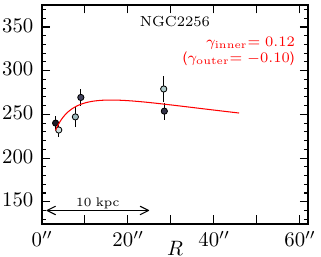}
\includegraphics[page=37]{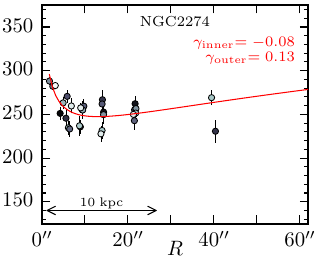}
\\
\includegraphics[page=1]{plotting/figs-allprofiles.pdf}
\includegraphics[page=38]{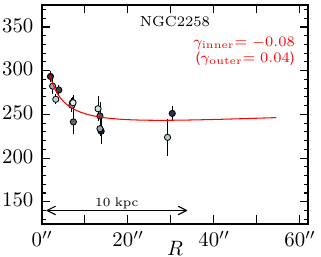}
\includegraphics[page=39]{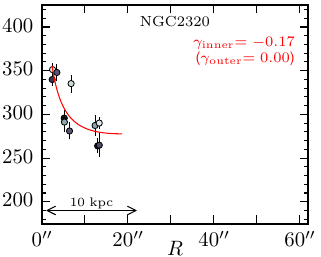}
\includegraphics[page=40]{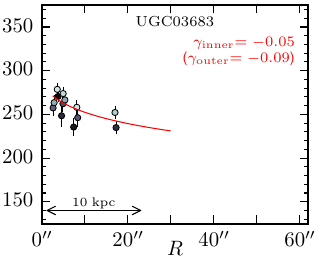}
\\
\includegraphics[page=1]{plotting/figs-allprofiles.pdf}
\includegraphics[page=41]{plotting/figs-allprofiles.pdf}
\includegraphics[page=42]{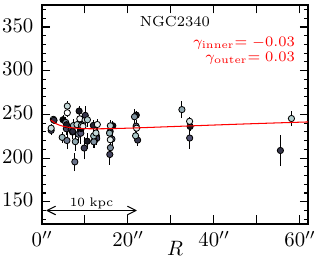}
\includegraphics[page=43]{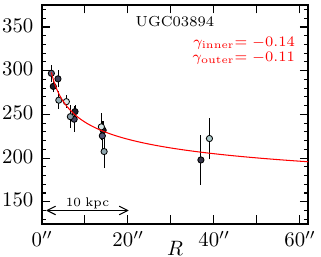}
\\
\includegraphics[page=1]{plotting/figs-allprofiles.pdf}
\includegraphics[page=44]{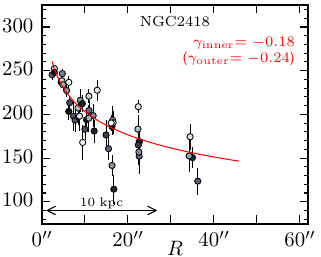}
\includegraphics[page=45]{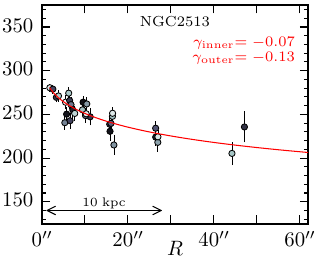}
\includegraphics[page=46]{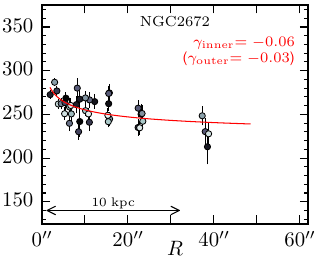}
\end{center}
\caption{(continued)}
\label{fig:allgals3}
\end{figure*}
\begin{figure*}
\begin{center}
\ContinuedFloat
\includegraphics[page=1]{plotting/figs-allprofiles.pdf}
\includegraphics[page=47]{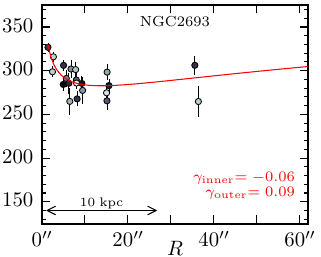}
\includegraphics[page=48]{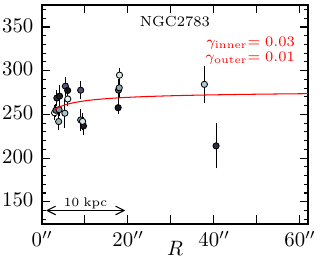}
\includegraphics[page=49]{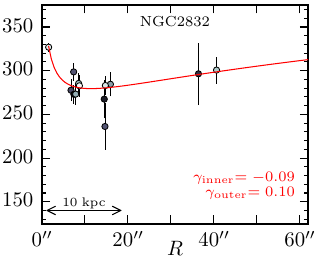}
\\
\includegraphics[page=1]{plotting/figs-allprofiles.pdf}
\includegraphics[page=50]{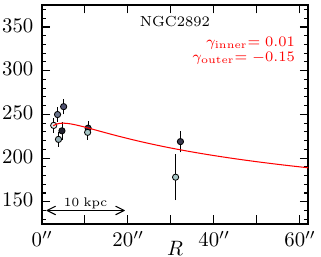}
\includegraphics[page=51]{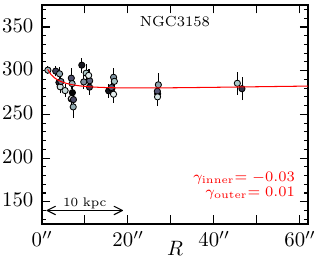}
\includegraphics[page=52]{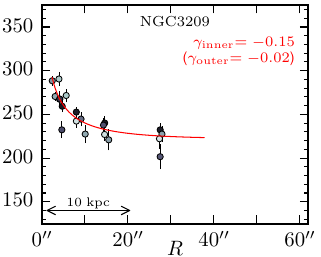}
\\
\includegraphics[page=1]{plotting/figs-allprofiles.pdf}
\includegraphics[page=53]{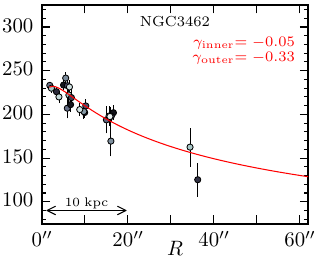}
\includegraphics[page=54]{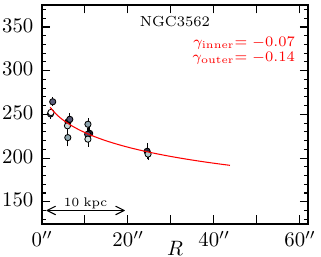}
\includegraphics[page=55]{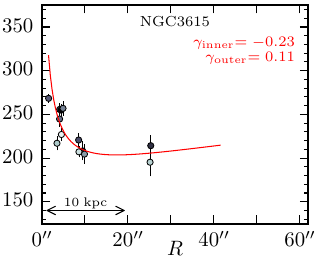}
\\
\includegraphics[page=1]{plotting/figs-allprofiles.pdf}
\includegraphics[page=56]{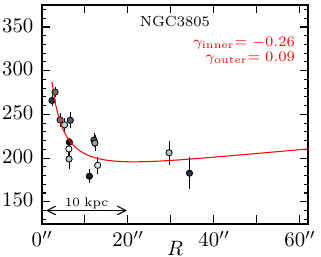}
\includegraphics[page=57]{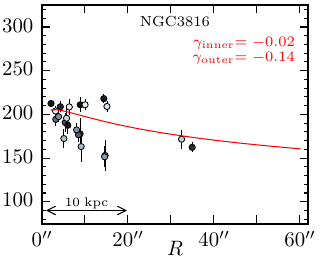}
\includegraphics[page=58]{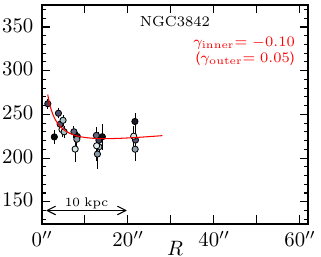}
\\
\includegraphics[page=1]{plotting/figs-allprofiles.pdf}
\includegraphics[page=59]{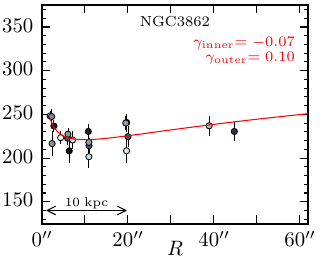}
\includegraphics[page=60]{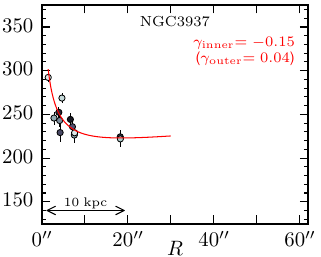}
\includegraphics[page=61]{plotting/figs-allprofiles.pdf}
\end{center}
\caption{(continued)}
\label{fig:allgals4}
\end{figure*}
\begin{figure*}
\begin{center}
\ContinuedFloat
\includegraphics[page=1]{plotting/figs-allprofiles.pdf}
\includegraphics[page=62]{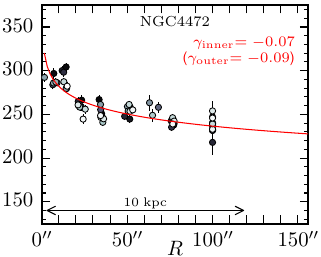}
\includegraphics[page=63]{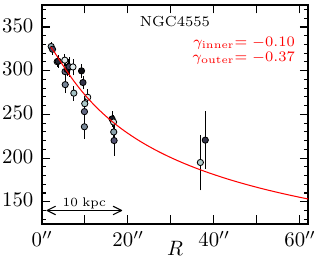}
\includegraphics[page=64]{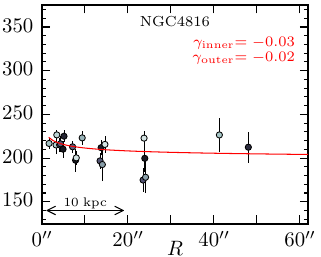}
\\
\includegraphics[page=1]{plotting/figs-allprofiles.pdf}
\includegraphics[page=65]{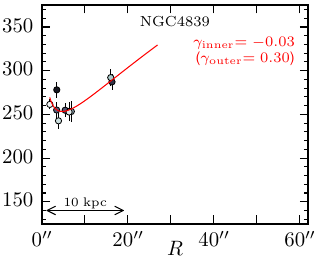}
\includegraphics[page=66]{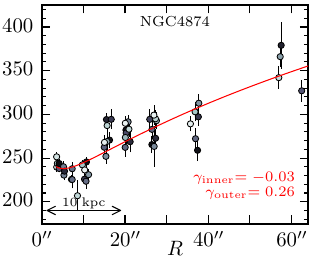}
\includegraphics[page=67]{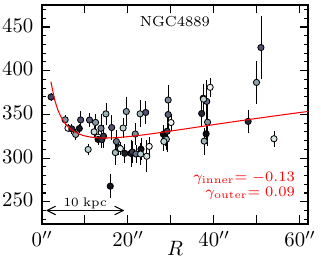}
\\
\includegraphics[page=1]{plotting/figs-allprofiles.pdf}
\includegraphics[page=68]{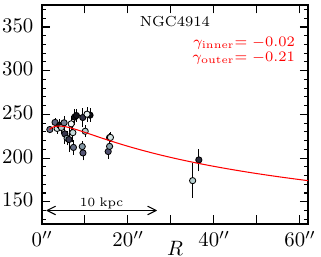}
\includegraphics[page=69]{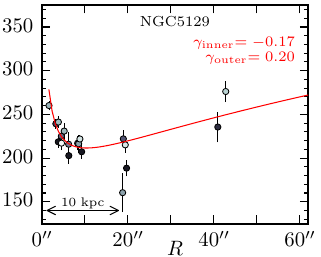}
\includegraphics[page=70]{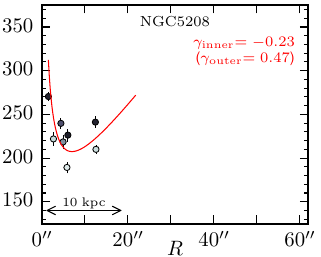}
\\
\includegraphics[page=1]{plotting/figs-allprofiles.pdf}
\includegraphics[page=71]{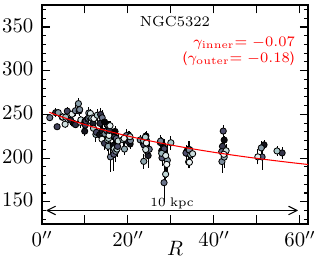}
\includegraphics[page=72]{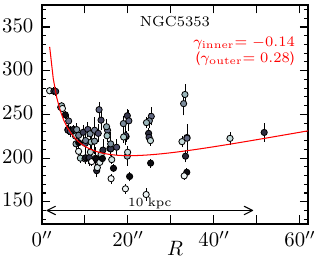}
\includegraphics[page=73]{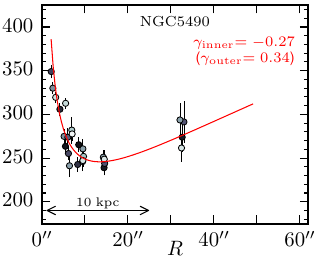}
\\
\includegraphics[page=1]{plotting/figs-allprofiles.pdf}
\includegraphics[page=74]{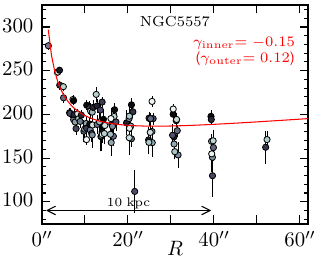}
\includegraphics[page=75]{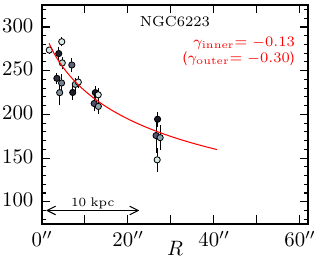}
\includegraphics[page=76]{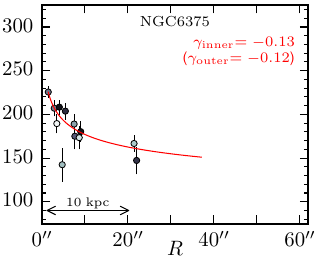}
\end{center}
\caption{(continued)}
\label{fig:allgals5}
\end{figure*}
\begin{figure*}
\begin{center}
\ContinuedFloat
\includegraphics[page=1]{plotting/figs-allprofiles.pdf}
\includegraphics[page=77]{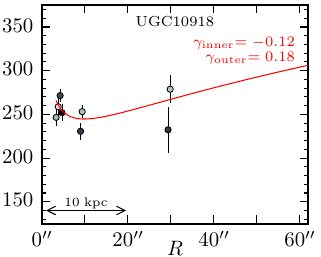}
\includegraphics[page=78]{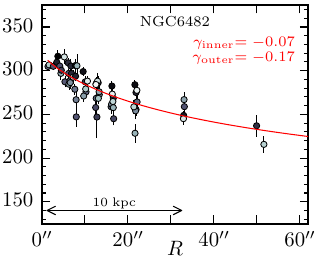}
\includegraphics[page=79]{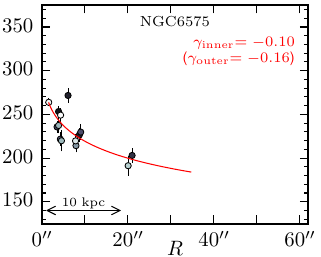}
\\
\includegraphics[page=1]{plotting/figs-allprofiles.pdf}
\includegraphics[page=80]{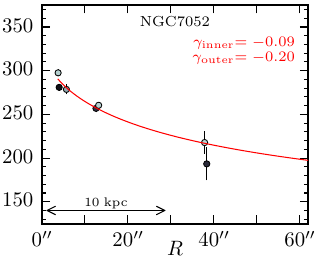}
\includegraphics[page=81]{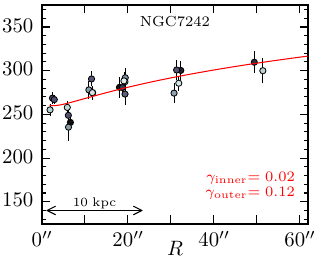}
\includegraphics[page=82]{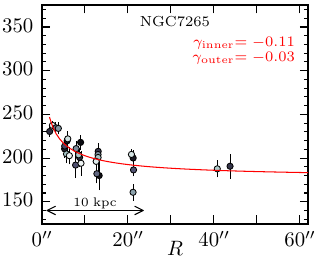}
\\
\includegraphics[page=1]{plotting/figs-allprofiles.pdf}
\includegraphics[page=83]{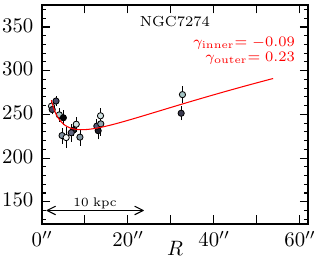}
\includegraphics[page=84]{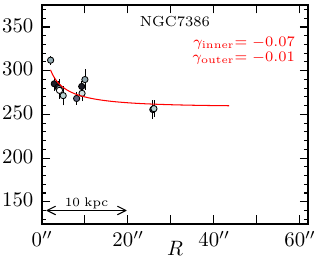}
\includegraphics[page=85]{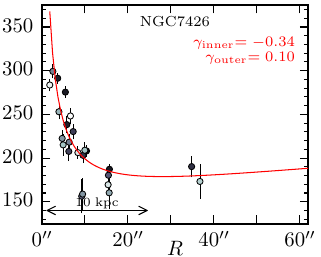}
\\
\includegraphics[page=1]{plotting/figs-allprofiles.pdf}
\includegraphics[page=86]{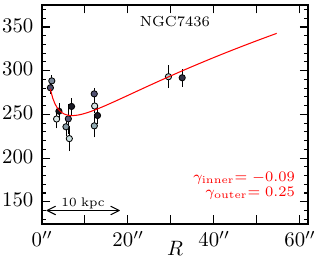}
\includegraphics[page=87]{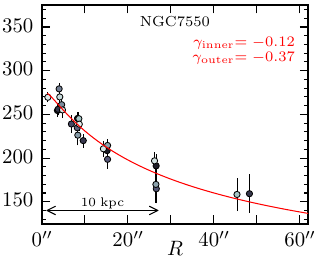}
\includegraphics[page=88]{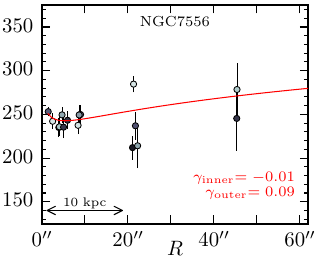}
\\
\includegraphics[page=1]{plotting/figs-allprofiles.pdf}
\includegraphics[page=89]{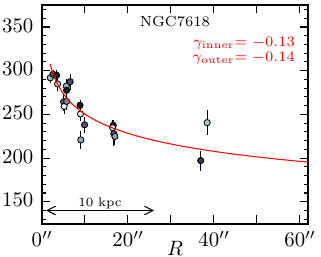}
\includegraphics[page=90]{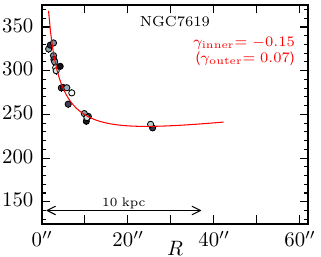}
\includegraphics[page=91]{plotting/figs-allprofiles.pdf}
\end{center}
\caption{(continued)}
\label{fig:allgals6}
\end{figure*}


\bsp	
\label{lastpage}
\end{document}